\documentclass[a4page]{article}
\usepackage{fullpage}
\usepackage{amsfonts}
\usepackage{mathtools} 
\usepackage{amsmath} 
\usepackage{amssymb} 
\usepackage[export]{adjustbox} 
\usepackage{multicol} 

\usepackage{color}
\usepackage{graphicx} 
\usepackage{float}
\usepackage[caption = false]{subfig}
\DeclareGraphicsExtensions{.pdf}
\graphicspath{ {./figures/} }

\title{Estimating information in time-varying signals}
\author{Sarah A Cepeda-Humerez, Jakob Ruess, Ga\v{s}per Tka\v{c}ik}
\usepackage{graphicx} 
\usepackage{etoolbox} 

\begin{document}
\maketitle

\begin{abstract}
Across diverse biological systems---ranging from neural networks to intracellular signaling and genetic regulatory networks---the information about changes in the environment is frequently encoded in the full temporal dynamics of the network nodes. A pressing data-analysis challenge has thus been to efficiently estimate the amount of information that these dynamics convey from  experimental data. Here we develop and evaluate decoding-based estimation methods to lower bound the mutual information about a finite set of inputs, encoded in single-cell high-dimensional  time series data. For biological reaction networks governed by the chemical Master equation, we derive model-based  information approximations and analytical upper bounds, against which we  benchmark our proposed model-free decoding estimators.  In contrast to the frequently-used k-nearest-neighbor estimator, decoding-based estimators robustly extract a large fraction of the available information from high-dimensional trajectories with a realistic number of  data samples. We apply these estimators to previously published data on Erk and Ca$^+$ signaling in mammalian cells and to yeast stress-response, and find that substantial amount of information about environmental state can be encoded by non-trivial response statistics even in stationary signals. We argue that these single-cell, decoding-based information estimates, rather than the commonly-used tests for significant differences between selected population response statistics, provide a proper and unbiased measure for the performance of biological signaling networks.
\end{abstract}

\section{Introduction}
For their survival, reproduction, and differentiation, cells depend on their ability to respond and adapt to continually changing environmental conditions. Environmental information must be sensed and often transduced to the nucleus, where an appropriate response is initiated, usually by selectively up- or down-regulating the expression levels of target genes. This information flow is mediated by biochemical reaction networks, in which concentrations of various signaling molecules encode for different environmental states or different response programs. This map between environmental input or response output and the internal chemical state is, however, highly stochastic, because typical networks operate with small absolute copy numbers of signaling molecules~\cite{Eldar2010}; moreover, different environments can be encoded by the same signaling molecule, by differentially regulating the dynamics of its concentration~\cite{Purvis2013}. This raises two fundamental questions: first, how much information the cells could, even in principle, encode in the combinatorial and possibly time-varying concentrations of multiple signaling molecules and how such information could be plausibly read out during ``downstream'' processing; and second, how can we quantify, in an unbiased and model-free fashion, the amount of information available to the cells from limited experimental data.

Information theory provides a framework within which the theoretical study of limits to communication as well as the empirical study of actual information flows  can be addressed~\cite{Shannon1949}. Applications of information theory to questions in biology and, in particular, neuroscience started already in the 1950s and continue to this day, with the main focus to understand how---and with what accuracy---neural activity encodes information about the environment~\cite{Paninski2003,Strong1998,quiroga2009extracting}. Applications of analogous techniques to biochemical signaling only started recently and represent an active area of research at the interface of physics, biology, statistics, and engineering~\cite{BOWSHER2014, BialekBook, RevGT, GTReview}. 

Recent theoretical work analyzed the reliability of information transmission through specific reaction systems in the presence of molecular noise, e.g., during ligand binding~\cite{Thomas2016}, in chemotaxis~\cite{Tostevin2009}, gene regulation~\cite{GT2008,sokolowski2015optimizing,sokolowski2016extending,tkacik2012optimizing,walczak2010optimizing,tkavcik2009optimizing,Rieckh2014}, biochemical signaling networks~\cite{Cheong2011}, etc., and asked how such transmission can be maximized by tuning the reaction rates. Generally, these studies focused on  steady state, by considering the information encoded in a single temporal snapshot of the reaction network at equilibrium given the input signals. Rigorous extensions to dynamical signals have been  either rare and only possible for simple cases, like the BIND channel~\cite{Thomas2016}, or required  specific operating regimes that permitted linearization and Gaussianity assumptions~\cite{Tostevin2009,tostevin2010mutual,deRonde2011}. At its core, the analysis of signal transduction through nonlinear noisy chemical systems requires one to have control over the distribution of concentration trajectories given the (possibly) time-varying inputs; even if it were possible to calculate this distribution in principle, the curse of dimensionality puts strong limits to the manipulations required to compute the information transmission. Consequently, problems of this kind are currently considered intractable in their full generality.

Empirical estimates of information transmission in biochemical networks similarly focused on the steady state~\cite{Dubuis2013, Voliotis2014}, or considered only specific, hand-picked dynamical features, such as the amplitude or the frequency of the response, as information carriers~\cite{Hansen2015}. Recent developments of fluorescent reporters and microfluidics have enabled the characterization of  dynamical responses at a single cell resolution using large ($>10^4$) numbers of sampled response trajectories, thereby permitting direct information estimates using generic estimators like the k-nearest-neighbors (knn)~\cite{Selimkhanov2014}. Existing approaches, however, suffer from severe limitations: they still require a prohibitive number of samples, especially when the response is distributed over multiple chemical species; or they necessitate uncontrolled assumptions about  trajectory features that are supposed to be ``relevant''. We recently proposed and applied decoding-based information estimators~\cite{Granados2018} as an alternative that draws on the past experiences in neuroscience~\cite{Borst1999,marre2015high,Rieke1993} to dissect the yeast stress-response network. In this paper we provide a detailed account of the new methodology, show that it alleviates the most pressing problems of existing approaches, and benchmark it against synthetic and real data.

\section{Models and Methods}
\subsection{Biochemical reaction networks}

At their core, cellular processes consist of networks of chemical reactions. A chemical reaction network consists of a set of $m$ molecular species $\{\tilde{X}_1,\tilde{X}_2,\ldots,\tilde{X}_m\}$ that interact through $K$ coupled reactions of the form:
\begin{equation}
\nu_{1k}'\tilde{X}_1+\ldots+\nu_{mk}'\tilde{X}_m \xrightarrow{\quad\theta_k\quad} \nu_{1k}''\tilde{X}_1+\ldots+\nu_{mk}''\tilde{X}_m , \quad k=1,\ldots,K
\label{eq:BiochemicalReactionNetwork}
\end{equation}
where $\nu_{1k}',\ldots,\nu_{mk}'$ and $\nu_{1k}'',\ldots,\nu_{mk}''$ are coefficients that determine how many molecules of each species are consumed and produced in the $k$-th reaction. $\theta_1\ldots\theta_k\in\mathbb{R}^+$ determine the rates at which the reactions occur and depend on binding affinities of chemical species, temperature and possibly the external conditions.

If we assume that the system is well-stirred, in thermal equilibrium and the reaction volume is constant, it can be shown that the probability that a reaction of type $k$ takes place in an infinitesimal time interval $[t,t+dt]$ can be written as $a_k(\tilde{x},\theta_k)dt=\theta_k g_k(\tilde{x})dt$, where $\tilde{x} = [\tilde{x}_1,\ldots,\tilde{x}_m]^T\in \mathbb{N}_0^m$ contains the amounts of molecules of the $m$ species that are present in the system at time $t$, and $g_k(\tilde{x})=\prod_{i=1}^m {\tilde{x}_i\choose\nu_{ik}'}$ counts all possibilities of choosing the required reaction molecules out of all available molecules~\cite{Gillespie1992,VanKampen2007}. $\theta_k$ is a  constant that depends on the physical characteristics of the cell but also on the environmental conditions. 

Let us denote the probability that $\tilde{x}$ molecules of the $m$ species are present in the system at time $t\in\mathbb{R^+}$ by $p(\tilde{x},t)$ and define the stoichiometric change vectors $\nu_k=[\nu_{1k},\ldots,\nu_{mk}]^T\in\mathbb{Z}^m,\; k=1,\ldots, K$, as the net changes in the amount of molecules in the reactions, i.e. $\nu_{ik}=\nu_{ik}''-\nu_{ik}',\, i=1,\ldots, m,\, k=1,\ldots, K$. Then it can be shown \cite{VanKampen2007} that the chemical master equation (CME) can be written as:
\begin{equation}
\dot{p}(\tilde{x},t)=-p(\tilde{x},t)\sum_{k=1}^{K}a_k(\tilde{x},\theta_k)+\sum_{k=1}^{K}p(\tilde{x}-\nu_k,t)a_k(\tilde{x}-\nu_k,\theta_k),
\end{equation}    
or in  a more compact form \cite{VanKampen2007}
\begin{equation}
\dot{\mathbf{p}}(t)= \mathbf{M} \mathbf{p}(t),
\label{eq:CME}
\end{equation}
where $\mathbf{p}(t)$ is a vector with components $p(\tilde{x},t)$, which is, in principle, infinite dimensional, and $\mathbf{M}$ contains the transition rates between all possible states, e.g. the transition rate from state $\tilde{x}_k'=\tilde{x}-\nu_k$ to state $\tilde{x}$ is given by
\begin{equation}
M_{\tilde{x},\tilde{x}_k'}=a_k(\tilde{x}_k',\theta_k)-\delta_{\tilde{x},\tilde{x}_k'}\sum_{q} a_q(\tilde{x},\theta_q),
\end{equation} 
where $\delta$ is the Kronecker delta, which is $1$ when $\tilde{x}=\tilde{x}_k'$ and $0$ otherwise.

The CME given in Eq~(\ref{eq:CME}) is an instance of a continuous-time discrete-state-space Markov Chain for a random process $X$ that can be solved exactly only for a few simple cases. It is nevertheless possible to efficiently generate samples $x$ of the random process $X$, which we will refer to as ``trajectories'' or ``paths'', for a selected time interval, $t\in[0,T]$, according to the correct probability distribution $p$, by the Stochastic Simulation Algorithm (SSA, or the Gillespie algorithm)~\cite{Gillespie1977}. 

To study information transmission through the biochemical networks described by the CME, we need to define the input and output signals. In the simplest setup considered here, the input $U$ is a discrete random variable that can take on one of the  $q\geq 2$ possible values, $U\in\{u^{(1)},u^{(2)},\dots,u^{(q)}\}$. Each  input in general corresponds to a distinct set of reaction rate constants $\theta$, but in models of real biological networks, changing input often modulates only one or a few rates in the system, e.g., by representing the change in a key external ligand concentration, receptor activity, etc. Changes in the input are reflected in changes in the resulting trajectories of chemical species amounts, $x$. Typically, only a subset of chemical species could be considered as biologically-relevant ``outputs'' that encode the information about the environmental change: this would correspond to marginalizing $p$ in Eq~(\ref{eq:CME}) over the unobserved (non-output) chemical species for the purposes of information transmission. While this is an interesting theoretical problem in its own right, here we work with simple toy examples where the output will be the trajectory, $x$, over the complete state space, i.e., we assume that all chemical species in the reaction network can be fully and perfectly observed. As we explain below, this allows us to define and compute the mutual information between a discrete input, $U$, and the output random process $X$ given by the CME in a straightforward fashion. We later show that this computation can be carried out also when the continuous-time process $X$ is sampled at uniform discrete times, as would be the case with experimental measurements.

\subsection{Mutual information between discrete inputs and response trajectories}
Information theory introduces the mutual information as the measure of fidelity by which changes in one random variable, e.g., the input $U$, can effect changes in another random variable, e.g., $X$. In this sense, mutual information is simply a measure of statistical dependency (i.e., any correlation, be it linear or not) between $U$ and $X$, and can thus be written as a functional of the joint probability density function $p(x,u)$:
\begin{equation}
I(X;U) = \int_X\int_U p(x,u)\log_2\left(\frac{p(x,u)}{p_X(x)p_U(u)}\right)\; du\,dx
\label{eq:informationContinuous}
\end{equation}
where $p_U$ and $p_X$ are the marginal density functions for $U$ and $X$, respectively, and we have generically written $u$ and $x$ as continuous variables; if they are discrete, integral signs are replaced by summations over the support for the corresponding probability distributions, as appropriate.

Mutual information is a non-negative symmetric quantity that is measured in bits, and is zero only if $X$ and $U$ are statistically independent. When studying information transmission through a channel $U\rightarrow X$ specified by $p(x|u)$, for which $U$ serve as inputs drawn from an input distribution $p_U(u)$, it is common to rewrite Eq.~(\ref{eq:informationContinuous}) as

\begin{equation}
I(X;U) = H(U) - H(U|X) = H(X) - H(X|U),
\label{eq:InfoEntropy}
\end{equation}
where $H(X)$ is the differential entropy of $X$ (and analogously for $H(U)$),  defined as 
\begin{equation}
H(X) = -\int_X p_X(x)\log_2 p_X(x)\; dx,
\label{eq:entropy}
\end{equation}
and the conditional entropy,  $H(X|U)$, is 
\begin{equation}
H(X|U) = \int_U H(X|u)p_U(u) \; du = -\int_U \int_X p_U(u)p(x|u)\log_2 p(x|u) \;dx \,du.
\end{equation} 

Equations~(\ref{eq:InfoEntropy}) can be interpreted in two ways: information is either the difference between the total variability in the repertoire of responses $X$ that the biochemical network can generate (measured by the \emph{response entropy}, $H(X)$) and the average variability at fixed input that is due to noise in the network (measured by the \emph{noise entropy}, $H(X|U)$); alternatively, information is also the entropy of the inputs, $H(U)$, minus equivocation $H(U|X)$, or the average uncertainty in what input was sent given that a particular response was observed. These interpretations make explicit the dependence of information both on the properties of the channel (the biochemical reaction network), as well as on the distribution of signals $p_U$ that the network receives. In this work, we will consider discrete inputs and will assume uniform $p_U$. It is, however, also possible to compute the \emph{channel capacity} $C$ by maximizing the information flow at given $p(x|u)$ over all possible input distributions,
\begin{equation}
C = \max_{p_U} \; I(X;U);
\end{equation}
Shannon's classic work then proves that error-free transmission at rates higher than those given by capacity is impossible, while error-free transmission at rates below capacity can be achieved with the optimal use of the channel. Contrary to engineering, where the focus is on finding encoding and decoding schemes that best utilize a given channel, in biophysics and systems biology mutual information is used as a tool to quantify the limits to biological signal processing due to noise without needing to make assumptions about possible biochemical encoding and decoding mechanisms. 

\begin{figure}[tb] 
	\centering
	\includegraphics[width=.8\textwidth]{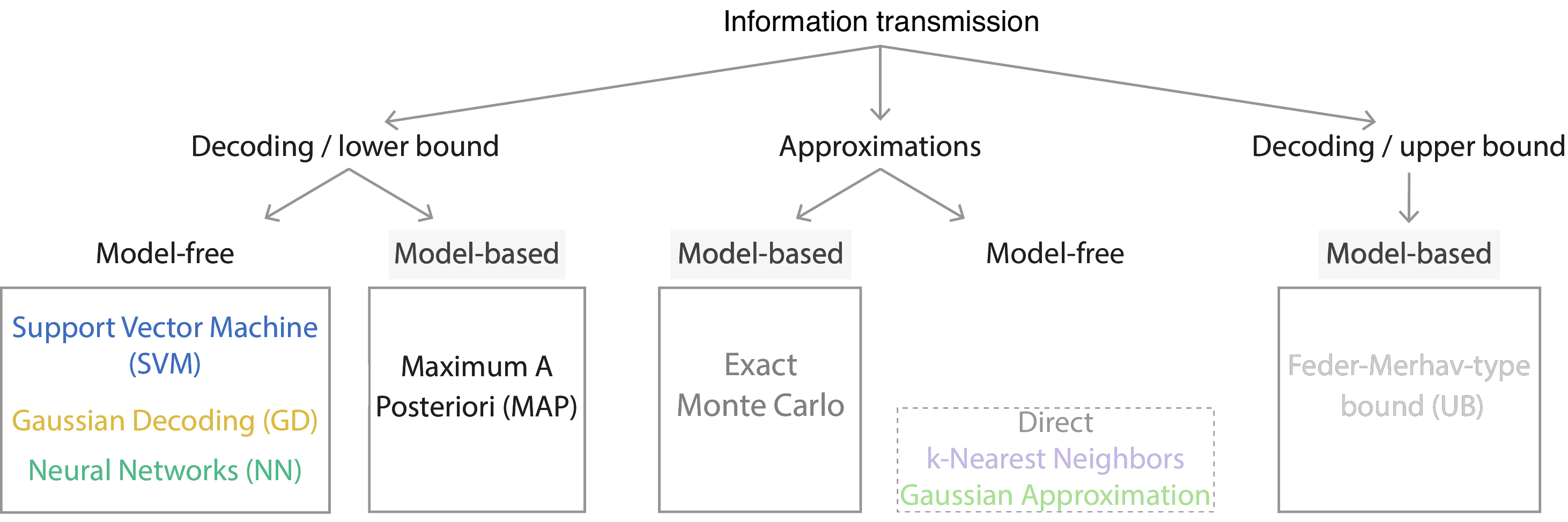}
	\caption{{\bf Information transmission between discrete inputs and response trajectories in biochemical networks.} For fully-observed reaction networks whose dynamics are governed by a known chemical Master equation, information can be approximated to an arbitrary accuracy via Monte Carlo integration for either continuous-time or discrete-time response trajectories (model-based \emph{exact Monte Carlo}, Section~\ref{sec:mexact}). Since full knowledge of the reaction system is used, these approximations are tractable deep in the regimes where model-free estimations break down with uncontrolled errors (Section~\ref{sec:mfe}). True information estimates are lower-bounded by model-based maximum a posteriori (MAP) or Bayes optimal decoding (Section~\ref{sec:lowerbound}). This decoding gives the lowest average probability of error and the corresponding information lower bound can be used as a benchmark for information estimates derived from other model-free decoding approaches (that have at least the error probability of the MAP decoder); in Section~\ref{sec:decest} we compare Support Vector Machine (SVM), Gaussian Decoding (GD) and Neural Network (NN) decoding approaches. Upper bounds like the Feder-Merhav bound~\cite{Feder1994} and our improvement on it~\cite{Hledik2018} complete the picture by estimating the gap between optimal decoding-derived and exact information values (Section~\ref{sec:lowerbound}).  } 
	\label{fig:scheme}	
\end{figure}

The setup we consider here is one in which inputs $U$ are iid drawn from a uniform distribution and change rarely, i.e., at a rate that is much lower than the (inverse) timescale on which the reaction network in Eq~(\ref{eq:BiochemicalReactionNetwork}) relaxes to its steady state. We assume that after an input change, we observe a fixed-time segment of the complete network dynamics, $x$, which is a sample path in $m$-dimensional discrete space, making direct calculation of information, $I(X;U)$, by  integrating / summing over all possible trajectories  as implied by Eq~(\ref{eq:informationContinuous}) intractable. We will nevertheless show that estimates of exact information are possible if the reaction network is known, by explicitly using the transition matrix $\mathbf{M}$ of the  Markov Chain from Eq~(\ref{eq:CME}) and generating exact sample paths, that is, realizations of $X$, using  SSA. We call this model-based approach \emph{exact Monte Carlo approximation} and contrast it to uncontrolled model-free estimations such as those obtained by using Gaussian approximations or k-nearest-neighbors methodology. We then introduce various decoding estimators and establish a hierarchy through which these estimates upper and lower-bound the true information, as shown in Fig~\ref{fig:scheme}.

\subsection{Exact information calculations for fully observed reaction networks}
\label{sec:mexact}
 {\bf Responses in continuous time.} Given the specification of the biochemical reaction network in Eq~(\ref{eq:BiochemicalReactionNetwork}), we sample $N$ trajectories, $x$, using the Gillespie (SSA) algorithm. Each trajectory $x$ can be represented as the sequence of consecutive states representing molecular species counts, $\textbf{s}=[s_{1},s_2,\ldots,s_r]$, where $s_1=\tilde{x}(t=0)$, etc., and $s_{r}=\tilde{x}(t=\sum_{i=1}^{r-1}t_i))$, and the sequence of  time intervals spent in each state, $\mathbf{t} = [t_1,\ldots,t_r],\;0<t_i<T,\; i=1,\ldots,r$ and $T = \sum_i^r t_i$. Then the likelihood of $x$ for a given input $u$ is: 
\begin{equation}
p(x|u) = p(s_1)\exp(M_{s_1s_1}t_1)\prod_{i=2}^{r}M_{s_is_{i-1}}\exp(M_{s_is_i}t_i)
\label{eq:likelihoodMarkovCont}
\end{equation} 
where $p(s_1)$ is given by the initial conditions of the process, and  the transition matrix $\mathbf{M}$ depends on the input $u$. To get the marginal distribution, $p_X(x)$, we sum over all possible input values:
\begin{equation}
 p_X(x) = \sum_{i=1}^{q} p({x}|u^{(i)})p_U(u^{(i)}).
 \label{eq:marginalpbbTrajectoryCont}
\end{equation}
Since we are able to compute the exact likelihood for each path generated by the stochastic process, entropic quantities can be approximated without significant biases using Monte Carlo integration, where the integral over states in Eq~(\ref{eq:entropy}) is replaced by an average over $N$ sampled trajectories:
\begin{equation}
\tilde{H}(X) =  - \frac{1}{N}\sum_{i = 1}^{N} \log_2 p_X({x_i}).
\label{eq:EntropyMonteCarloApproximation}
\end{equation}
Similarly, we can approximate $H(X|U)$:
\begin{equation}
\tilde{H}(X|U) =- \sum_{j=1}^{q}p_U(u^{(j)})\left(\frac{1}{N}\sum_{i=1}^N \log_2 p({x_i}|u^{(j)})\right).
\label{eq:conditionalentropyMC}
\end{equation}
The exact Monte Carlo information approximation is finally obtained using Eq~(\ref{eq:InfoEntropy}):
\begin{equation}
I_{\rm exact}^{\rm *} = \tilde{H}(X)-\tilde{H}(X|U),
\label{eq:Itrue}
\end{equation}
where $*$ reminds us that the paths are represented in continuous time.\\

\noindent {\bf Responses in discrete time.} 
We can resample the continuous trajectories $X$ on a grid of uniformly spaced time points to obtain a new discrete random variable, $\textbf{X}=[X(t=\Delta t),\ldots,X(t=i\Delta t).\ldots,X(t=d\Delta t)]\in \mathbb{N}^{m\times d}_0$, where $\Delta t$ is the discretization step, $d=T/\Delta t$ is the length of $\mathbf{X}$. For convenient notation we denote this random variable as $\textbf{X}=[X^0,\ldots, X^d]$, and its realizations, the discrete trajectories, as $\textbf{x}$.

In the discrete case, the likelihood of $\textbf{x}$ for a given input $u$ can be computed using the general solution to Eq~(\ref{eq:CME}):
\begin{equation}
\mathbf{p}(t) = e^{\mathbf{M}t}\mathbf{p}(0), \label{eq:mx}
\end{equation}
where $\mathbf{p}(t)$ is the probability distribution of states after time $t$, with the initial probability distribution $\mathbf{p}(t=0)=\mathbf{p}(0)$. Using this formal solution we compute the transition matrix between discrete timesteps separated by $\Delta t$ to get:
\begin{equation}
\mathbf{W}=e^{\mathbf{M}\Delta t},
\end{equation}
where $\mathbf{M}$ and thus $\mathbf{W}$ again depend on $u$. 
The likelihood of any discrete path can then be obtained by multiplying the transition probabilities between all the $d$ consecutive states in the path for a given input $u$:
\begin{equation}
p(\mathbf{x}|u) = p(\mathbf{x}^0)\prod_{i=2}^{d}\mathbf{W}_{\mathbf{x}^i \mathbf{x}^{i-1}}.
\label{eq:likelihoodMarkovDiscrete}
\end{equation}
We can now approximate the information between input $U$ and a discretely sampled trajectory $\mathbf{X}$, $I_{\rm exact}$, as in the continuous case: we get the marginal $p_X(\mathbf{x})$ with Eq~(\ref{eq:marginalpbbTrajectoryCont}) and use Eqs~(\ref{eq:EntropyMonteCarloApproximation}, \ref{eq:conditionalentropyMC}) in Eq~(\ref{eq:Itrue}). In general, temporal discretization loses information relative to the full (continuous-time) trajectory, where reaction events in the trajectory $x$ are recorded with infinite temporal precision, so the information in discretely-sampled trajectories, $I$, must be bounded from above by the information in continuous-time trajectories, $I^*$:
\begin{equation}
I_{\rm exact}(\mathbf{X};U) \leq I_{\rm exact}^*(X;U),
\label{eq:ContDiscrIneq}
\end{equation}
where equality is approached in the limit of ever finer temporal discretization, $\Delta t\rightarrow 0$.
\subsection{Model-free information estimators}
\label{sec:mfe}
In the absence of a full stochastic model for the biochemical reaction network, mutual information estimation is tractable only if we make assumptions about the distribution of response trajectories given the input. We briefly summarize two approaches below: in the first, k-nearest-neighbor procedure, the space in which the response trajectories are embedded is assumed to have a particular metric; in the second, Gaussian approximation, we assume a particularly tractable functional form for the channel, $p(\mathbf{x}|u)$. 
\\

\noindent {\bf K-nearest-neighbors (knn) estimator.} 
The idea of using the nearest neighbour statistics to estimates entropies is at least 70 years old~\cite{Dobrushin1958,Vasicek1973}, while estimators for mutual information have been developed during the early 2000s~\cite{Kaiser2002,Kraskov2004}. The cornerstone of the approach is to compute the estimate from the distances of $d$-dimensional real valued data points to their $k$-th nearest neighbour.  Hence, the estimator depends on the metric chosen to define this distance. Furthermore, its performance is known to depend on the value of $k$ (number of nearest neighbours), where small $k$ increase the variance and decrease the bias~ \cite{Khan2007}. This method has been used in several studies that estimated mutual information from single cell time series~\cite{Potter2017, Selimkhanov2014,Voliotis2014}. These studies used large numbers of response trajectories to provide the first evidence that the information available from the full timeseries of the response could be substantially higher than the information available from any response snapshot. 
\\

\noindent {\bf Gaussian approximation.}
A simplifying assumption in the Gaussian approximation is that the distribution of trajectories sampled at discrete times given input is approximately Gaussian, with the mean $\mathbf{\mu} \in \mathbb{R}^d$ and covariance matrix $\mathbf{\Sigma} \in \mathbb{R}^{d\times d}$ that may both depend on the input, $u$:   
\begin{equation}
p(\mathbf{x}|u)=\mathcal{N}(\mathbf{x}; \mathbf{\mu}(u),\mathbf{\Sigma}(u))=\frac{1}{\sqrt{\det(2\pi \mathbf{\Sigma})}}\exp\left(-\frac{1}{2} (\mathbf{x}- \mathbf{\mu})^T \mathbf{\Sigma}^{-1} (\mathbf{x}- \mathbf{\mu})\right ). \label{eq:gc}
\end{equation}
The entropy of the multivariate distribution in Eq~(\ref{eq:gc}) has an analytical expression that only depends on $\mathbf{\Sigma}$:
\begin{equation}
H_{\rm G}(\mathbf{X}|u) = \frac{1}{2}\log(\det(2\pi e \mathbf{\Sigma}(u))),
\label{eq:G_entropy}
\end{equation} 
which can be averaged over $p_U(u)$ to get the conditional entropy, $H_G(\mathbf{X}|U)$.
To estimate the information, we further need $H(\mathbf{X})$ from Eq~(\ref{eq:InfoEntropy}). This  entropy of a Gaussian mixture has no closed form solution, but can be computed  by Monte Carlo integration as in the previous section, following discrete analogs of Eqs~(\ref{eq:marginalpbbTrajectoryCont},\ref{eq:EntropyMonteCarloApproximation}): we draw  random samples from each of the $q$ different multivariate Gaussian distributions, Eq~(\ref{eq:gc}), one for each possible input $u$, and assign the marginal probabilities to each sample $\mathbf{x}$ as
\begin{equation}
p_X(\mathbf{x}) = \sum_{i=1}^{q} p_U(u^{(i)})\mathcal{N}(\mathbf{x}; \mu(u^{(i)}),\mathbf{\Sigma}(u^{(i)})),
\end{equation}
permitting us to use Eq~(\ref{eq:EntropyMonteCarloApproximation}) to approximate the total entropy of output trajectories in the Gaussian approximation, $H_G(\mathbf{X})$, and thus to obtain the Gaussian estimate for the information, $I_{\rm G}(\mathbf{X};U) = H_{\rm G}(\mathbf{X}) - H_{\rm G}(\mathbf{X}|U)$. 

To apply this estimator, one must use real (or simulated) data to estimate the conditional mean, $\mathbf{\mu}(u)$, and conditional covariance, $\mathbf{\Sigma}(u)$ for every possible $u$, from a limited number of samples. While general caveats for such estimations have been detailed in many textbooks~\cite{anderson1958introduction}, we emphasize that information estimation is particularly sensitive due to the computation of the determinant in Eq~(\ref{eq:G_entropy}) which can easily lead to ill-posed numerics when the number of samples is small. This can be mitigated by various regularization methods (one of which, the diagonal regularization, we  demonstrate later) that impose a prior structure on the estimated covariance. Yet even in the case of significant oversampling that we can explore using simulated data, the Gaussian approximation introduced here---in contrast to Gaussian decoding estimator introduced in the next section---can provide information values that deviate significantly from the true value and are not guaranteed to bound the true value from either above or below. This is because the true solutions of the CME live in the positive quadrant of the discrete space, and are thus essentially different from the Gaussian distributions assumed here. We nevertheless present this estimator because (i) it forms the basis for the Gaussian decoding estimator, introduced below, and (ii) real data itself often deviates from stochastic trajectories sampled from the CME in that it is continuous (since we measure, e.g., fluorescence proxy for a concentration of a protein of interest) and contains extra noise, making Gaussian approximation potentially applicable.

\subsection{Decoding-based information bounds}
 \label{sec:lowerbound}
 Here and in the next section we introduce a class of decoding-based calculations that lower-bound the exact information, $I(\mathbf{X};U)$, and can tractably be used as information estimators over realistically-sized data sets.
  Let  $\mathcal{D}$ consist of a set of $N$ labeled paths, typically represented in discretely sampled time, $\mathcal{D}=\{(u_1,\mathbf{x}_1),(u_2,\mathbf{x}_2),\dots,(u_{
 \rm N},\mathbf{x}_{\rm N})\}$, where $u_i$ and $\mathbf{x_i}$, for $i=1,\dots,N$, are realizations of the random variables $U\in \{u^{(1)},\dots,u^{(q)}\}$ and $\mathbf{X}\in \mathbb{R}^{m\times d}$, respectively. Here, $\mathcal{D}$ can represent either real data (typically containing $N \sim10^2-10^3$ trajectories) in case of model-free information estimates, or trajectories generated by exact simulation algorithms (in which case the sample size, $N$, is not limiting) from the full specification of the biochemical reaction network in case of model-based approximations. 
 
The procedure of estimating the input $\hat{u}$ from $\mathbf{x}$, such that the estimated $\hat{u}$ is ``as close as possible'' to true $u$ for a given trajectory $\mathbf{x}$, is known as decoding in information theory and neuroscience, and can equivalently be viewed as a classification task in machine learning or as an inference task in statistics. This procedure is implemented by a decoding function,
\begin{equation}
\hat{u} = F_\omega(\mathbf{x});
\end{equation}
$F$ is typically parametrized by parameters $\omega$ that need to be learned from data for model-free approaches, or derived from biochemical reaction network specification in case of model-based approaches.  $F$ assigns to every $\mathbf{x}_i$ in the dataset a corresponding ``decode'' $\hat{u}_i$ from the same space over which the random variable $U$ is defined; formally, these decodes are instances of a new random variable $\hat{U}$. The key idea of using decoding for information estimation starts with the observation that random variables 
 \begin{equation}
U \rightarrow X \xrightarrow{\mathrm{T_d}}  \mathbf{X} \xrightarrow{ F_{\omega}} \hat{U},
 \end{equation}
where $T_d$ represents time discretization, form a Markov chain. In other words, the distribution of $\hat{U}$ is conditionally independent of $U$ and only depends on $\mathbf{X}$, $p(\hat{u}|\mathbf{x},u) = p(\hat{u}|\mathbf{x})$, and so
\begin{equation}
p(\hat{u},\mathbf{x},u) = p_U(u)p(\mathbf{x}|u)p(\hat{u}|\mathbf{x}).
\end{equation}
The data processing inequality~\cite{Cover2005} can be used to further extend the bounds in Eq~(\ref{eq:ContDiscrIneq}): 
\begin{equation}
I(U;\hat{U})\leq I_{\rm exact}(U;\mathbf{X}) \leq I^*_{\rm exact}(U;X), \label{eq:finaldpi}
\end{equation}
where equality between the fist two terms holds only if $I(U;\mathbf{X}|\hat{U})=0$. Consequently, $I(U;\hat{U})$ is a lower bound to the information between trajectories $X$ and the input $U$~\cite{Brunel1998}. Note that analogous reasoning holds for decoding directly from continuous-time trajectories $X$. Better decoders which increase the correspondence between the true inputs and the corresponding decoded inputs will typically provide a tighter lower bound on the information. 

To compute the information lower bound, we apply the decoding function to each trajectory in $\mathcal{D}$ in model-based approximations or to each trajectory in the testing dataset for model-free estimators that need to be learned over training data first. We subsequently construct a $q\times q$ confusion matrix, also known as an error matrix, where each element $\epsilon_{ij}$ counts the fraction of realizations of $\mathbf{x}$ generated by an input $u=u^{(i)}$ that decode into $\hat{u}=u^{(j)}$. This matrix provides an empirical estimate of the probability distribution ${p}(\hat{u}, u)$, which can thus be used to compute the information estimate:
\begin{equation}
I(\hat{U};U) = \sum_{u,\hat{u}}  p(\hat{u},u)\log_2 \frac{p(\hat{u},u)}{p_U(u)p_{\hat{U}}(\hat{u})}  \approx \sum_{i=1}^{q}\sum_{j=1}^{q} \epsilon_{ij} \log_2 \frac{ \epsilon_{ij}}{\left(\sum_{k} \epsilon_{kj}\right)\left( \sum_{l} \epsilon_{il}\right)} , \label{eq:errorinfo}
\end{equation}
Crucially, in this estimation $O(N)$ data points are used to empirically estimate the elements of a $q \times q$ matrix $\epsilon$, and information estimation involves a tractable summation over these matrix elements; in contrast, direct estimates of $I(U;\mathbf{X})$ would involve an intractable summation over (vastly undersampled) space for $\mathbf{X}$. For typical applications where $q$ is small, decoding thus provides an essential dimensionality reduction prior to information estimation: in a simple but biologically relevant case of two distinct stimuli ($q=2$), information estimation only requires us to empirically construct a $2\times 2$ confusion matrix. If required, one can apply well-known debiasing techniques for larger $q$~\cite{Strong1998}. \\

\noindent {\bf Maximum a posteriori (MAP) lower bound.} In MAP lower bound, the decoding function $F_\omega$ is given by Bayesian inference of the most likely input $u$ given that a response trajectory $\mathbf{x}$ was observed, under the exact probabilistic model for the biochemical reaction network. MAP decoder is optimal in that it provides the lowest average probability of error, $\mathrm{Pr}(\hat{U}\neq U)$, among all decoders.  Typically, this will lead to a high mutual information value $I(\hat{U};U)$ compared to other (sub-optimal) decoders whose probability of error will likely be higher, making the information lower bound from MAP decoder a good benchmark for other decoder-based information estimates. We remind the reader, however, that even though MAP decoder achieves minimal error and typically high $I(\hat{U};U)$ values, this does not mathematically guarantee that its information will \emph{always} be higher or equal to the information of any other possible decoder, a fact that can be demonstrated explicitly using toy examples.

The MAP inference consists of finding the input that maximizes the posterior distribution~\cite{Murphy2012}
\begin{equation}
p(u|\mathbf{x}) = \frac{p(\mathbf{x},u)}{p_\mathbf{X}({\mathbf{x}})}= \frac{p(\mathbf{x}|u)p_U(u)}{p_\mathbf{X}({\mathbf{x}})}.
\end{equation}
This corresponds to the following decoding function:
\begin{equation}
\hat{u} = F_\omega(\mathbf{x}) = \mathrm{argmax}_{u} \left[ \log p(\mathbf{x}|u) + \log p_U(u)\right ],
\label{eq:mapdecoder}
\end{equation}
where $\omega$ represents the specification of the biochemical reaction network which determines $p(\mathbf{x}|u)$.
Here, $p_U(u)$ is assumed to be known, and the likelihood $p(\mathbf{x}|u)$ can be calculated using Eqs.~(\ref{eq:likelihoodMarkovCont}) or (\ref{eq:likelihoodMarkovDiscrete}), for the continuous or discrete time representations, respectively. 

One can apply the MAP-decoding based calculation of $I_{\rm MAP}(\hat{U};U)$ in two ways. First, when applied over real data $\mathcal{D}$, one can think of the procedure as a proper statistical estimation assuming that the biochemical network model is the correct generative model of the data (with estimation bias arising if it is not). Second, when applied, as we will do in the Results section, over trajectories $\mathcal{D}$ generated using exact stochastic simulation from the biochemical network model in the large $N$ limit, this procedure is a Monte Carlo approximation to the information lower bound.

Note that even though the MAP decoder is optimal, it does not follow that $I_{\rm MAP}(\hat{U};U) = I(\mathbf{X};U)$. This is because optimal channel use that realizes $I(\mathbf{X};U)$ may need to employ block codes, where a \emph{sequence} of inputs is encoded jointly into a \emph{sequence} of trajectories, which is later also jointly decoded. In contrast, the decoding bound $I_{\rm MAP}(\hat{U};U)$ relies on one-shot use of the channel: a single input $u$ is transduced into $\mathbf{x}$ which can immediately be decoded back into the estimate of the input, $\hat{u}$, on the basis of which the cell might make a decision. For many biological situations, this decoding setup should be more appropriate than the exact information calculation, as cells often need to react to stimuli as rapidly as possible in order to gain a selective advantage. Furthermore, it is difficult to conceive of biologically realistic encoders that would transform inputs into a block code in order to use the biochemical network channels optimally. \\

\noindent {\bf Maximum a posteriori upper bound (UB).}
\label{methods:fm}
Given that the optimal MAP decoding does not necessarily reach the exact mutual information, it is reasonable to ask how large the gap is between these two quantities. For discrete inputs, classic work in information theory proved a number of upper bounds on this gap when the channel is known~\cite{Samengo2002}, with the Feder-Merhav bound perhaps being the most well known~\cite{Feder1994}; Feder-Merhav provides an upper bound on the channel capacity given the overall probability of error in MAP decoding. In a separate work~\cite{Hledik2018}, we computed a  new upper bound on information $I_{\rm UB}(U;\mathbf{X})$ that is consistent with not just the overall probability of error as in Feder-Merhav bound, but with the full confusion matrix $\epsilon$ obtained from optimal MAP decoding, and showed that the new bound is tight.
 
Our self-contained derivation~\cite{Hledik2018} gives the following result
\begin{equation}
I(U;\mathbf{X}) \leq I_{\rm UB} = H(U) - \sum_{\hat{u}}p_{\hat{U}}(\hat{u}) \phi(\pi_{\hat{u}}),
\label{eq:UpperBound}
\end{equation}
where $\pi_{\hat{u}}=\mathrm{Pr}(U\neq\hat{U}|\hat{u})=1-\mathrm{Pr}(U=\hat{u}|\hat{U}=\hat{u})$ and functions $\phi$ and $\alpha$ can be expressed with the help of the floor and ceiling functions as:
\begin{align}
\phi(\pi)&=\alpha(\pi)\log_2\left \lfloor \frac{1}{1-\pi}\right\rfloor +(1-\alpha(\pi)) \log_2 \left \lceil \frac{1}{1-\pi}\right \rceil \label{eq:phi} \\
\alpha(\pi) &= \left \lfloor \frac{1}{1-\pi}\right\rfloor \bigg((1-\pi) \left \lceil \frac{1}{1-\pi}\right \rceil -1\bigg).
\label{eq:alpha}
\end{align}
This bound applies irrespectively of how the response trajectory space is represented (continuous or discrete, possibly of dimensionality much larger than that of the random variable $U$), since it is stated solely in terms of the input variable $U$ and its MAP decode, $\hat{U}$.
\\

\subsection {Decoding-based information estimators}
\label{sec:decest}

\noindent {\bf Support Vector Machine (SVM) lower bound estimator.} 
The first model-free decoding approach we consider is based on  classifiers called Support Vector Machines (SVMs). To begin we consider two possible inputs, $q=2$. We define a decoding function $F_\omega$ by means of a helper function $f_\omega(\mathbf{x})$, such that $F_\omega(\mathbf{x}) = u^{(1)}$ if $\mathrm{sign} f_\omega(\mathbf{x}) = -1$ and $F_\omega(\mathbf{x}) = u^{(2)}$ otherwise. Here, 
\begin{equation}
f_\omega(\mathbf{x}) = \sum_{i=1}^{N_{\rm t}}\alpha_i k (\mathbf{x}_i,\mathbf{x}) +b
\end{equation}
where $k:\mathbb{R}^d\times \mathbb{R}^d\rightarrow \mathbb{R}$ is the so-called ``kernel function'' to be defined below, $b$ is the bias constant, $N_{\rm t}$ is the number of samples in $\mathcal{D}_{\rm train}$ and $\alpha_1,\dots,\alpha_{N_{\rm t}}$ are obtained by solving standard SVM equations:
\begin{equation}
\min_{\substack{\alpha_1,\dots,\alpha_{N_{\rm t}} \in \mathbb{R},\\  \xi_1,\dots,\xi_{N_{\rm t}}\in \mathbb{R^+}}}\sum_{i,j=1}^{N_{\rm t}}\alpha_i \alpha_j k (x_i,x_j) + \frac{C}{N_{\rm t}} \sum_{i=1}^{N_{\rm t}} \xi_i
\end{equation}
subject to 
\begin{equation}
y_i\sum_{j=1}^{N_{\rm t}}\alpha_i k (\mathbf{x}_j,\mathbf{x}_i)\geq 1- \xi_i, \qquad \text{for} \enspace i=1,\dots,N_{\rm t}.
\end{equation}
$y_i=-1$ whenever the input corresponding to the $i$-th trajectory in the training set, $\mathbf{x}_i$, is $u^{(1)}$, i.e., $u_i=u^{(1)}$; similarly $y_i=+1$ whenever the corresponding input is $u^{(2)}$, i.e., $u_i=u^{(2)}$. $C$ is a positive regularization constant. Together, the parameters of the decoding function are $\omega=\{b,\mathbf{\alpha},\mathbf{\xi}, C\}$.

To prevent overfitting and set the regularization parameter $C$ using cross-validation, we split the full dataset $\mathcal{D}$ into training data, $\mathcal{D}_{\rm train}$, that consists of $N_t$ (here $
\sim 70\%$ of the total, $N$) of labeled sample trajectories, chosen randomly but balanced across different inputs $u$; the remaining $30\%$ of the data constitutes the testing data, $\mathcal{D}_{\rm test}$. Parameters $\omega$ are estimated only over $\mathcal{D}_{\rm train}$, after which the error matrix $\epsilon$ and the corresponding information estimate $I_{\rm SVM}(\hat{U}; U)$ of Eq~(\ref{eq:errorinfo}) are evaluated solely over $\mathcal{D}_{\rm test}$. The test/train split procedure can be repeated multiple times to compute the mean and the bootstrapped error bar estimate for the information estimator, $I_{\rm SVM}$~\cite{Granados2018}.

When we apply SVM decoding, we are still free to choose the kernel function. Here, we focus on two possibilities: 
\begin{itemize}
\item {\bf Linear kernel,} $k(x,x') = x^Tx'.$ The information estimate is based on a linear classifier that can learn to distinguish responses that differ in their conditional means, $\mu(u)$, but will result in close to chance performance if they don't. This is the simplest model-free decoding estimator and is thus a useful benchmark for more complex, non-linear decoders.
\item {\bf Radial basis functions kernel,} $k(x,x') = \exp(-\frac{||x-x'||^2}{2\sigma^2})$. This model-free decoder can be sensitive both to difference in the conditional means as well as higher-order statistics, e.g., the covariance matrix. Parameter $\sigma$ is set via cross-validation to maximize the performance.
\end{itemize}

For multiclass classification we use a decision-tree SVM classification method~\cite{Bennani2006}, also called Dendrogram-SVM (DSVM)~\cite{Lajnef2015}. To translate the multi-class classification into the canonical binary classification problem, this  method uses hierarchical bottom-up clustering to define the structure of the graph, on which a binary classification is performed using SVMs at each graph node. \\ 
 
 \noindent {\bf Gaussian decoder (GD) lower bound estimator.}
In this model-free estimation, we revisit the assumption that the (discretely sampled) output trajectories $\mathbf{x}$ given input $u$ can be approximated with a multivariate Gaussian distribution,  Eq.~(\ref{eq:gc}). The decoding function is then
\begin{equation}
\hat{u} = F_\omega(\mathbf{x})= \textrm{argmax}_{u} \left[\log \mathcal{N}\left({\bf x}; \mu(u),\mathbf{\Sigma}(u)\right) + \log p_U(u) \right]. \label{eq:gd}
\end{equation}
Here, parameters $\omega$ consist of conditional means and (possibly regularized) covariance matrices of the Gaussian distributions that need to be estimated from data, following the test/train procedure analogous to SVM decoding.

This method can be used with different parametric multivariate probability density functions replacing the multivariate Gaussian in Eq~(\ref{eq:gd}), with choices that approximate the statistics of the data (and thus the CME-derived distribution) better providing tighter lower bound estimates, $I_{\rm GD}(\hat{U};U)$, of the exact information.  By analogy with the exact MAP decoding using CME-derived response distribution, this method can also be understood as maximum a posteriori decoder but using approximate response distributions that need to be estimated from data. Here we decided to use the Gaussian distributions because they are the most unstructured (random) distributions based on measured first- and second-order statistics of the data. GD decoder thus should be able to discriminate various inputs if their responses differs either in the response mean or response covariance.\\

\noindent {\bf Neural Network (NN) lower bound estimator.}

Artificial neural networks, first introduced by the neurophysiologist Waren McCulloch and the mathematician Walter Pitts in 1943~\cite{McCulloch1943}, are nowadays the method of choice for classification that generally outperforms alternative machine learning techniques on very large and complex problems. Here we use one of the simplest neural networks, called the multi-layer perceptron (MLP). MLP is composed of layers of linear-threshold units (or LTUs), where each LTU computes a weighted sum of its inputs $z=\mathbf{\omega}^T \mathbf{x}$, then applies an activation function to that sum and outputs the result $y=h(z)=h(\mathbf{\omega}^T \mathbf{x} + \omega_0)$. Using a single LTU amounts to training a binary linear classifier by learning the weights $\omega$. As with linear SVM, such classifier only has a limited expressive power~\cite{Rosenblatt1957}, which can, however, be extended by stacking layers of LTUs so that outputs of the first layer are inputs to the second layer etc. 

For illustrative purposes we choose for our decoding function $F_\omega(\mathbf{x})$ a fully connected neural network with two hidden layers (with 300 and 200 LTUs, respectively) that uses the exponential activation function with $\alpha=1$:

	\[
	h_{\alpha}(z) =
	\begin{cases}
	\alpha (\exp(z)-1) & \text{if $z\leq0$} \\
	z & \text{if $z> 0$} 
	\end{cases}.
	\]

For training, we used He-initialization, which initializes the weights with a random number from a normal distribution with zero mean and standard deviation $\sigma = 2/ \sqrt{n_{in}}$, where $n_{in}$ is the number of inputs to units in a particular layer~\cite{Geron2017}, and Adam optimization with batch normalization and drop-out regularization~\cite{tensorflow2015,Geron2017}.  As before, we trained the neural network on  $\mathcal{D}_{\rm train}$, followed by the evaluation of the error matrix $\epsilon$  and of the corresponding information estimate, $I_{NN}(\hat{U};U)$, from Eq~(\ref{eq:errorinfo}), over $\mathcal{D}_{\rm test}$. We emphasize that the detailed architecture of the neural network we selected here is not relevant for other estimation cases; in general, the architecture is completely adjustable to the problem at hand and should be selected depending on the size of the training dataset. The only selection criterion is the network performance on test data, with better performing networks for a given dataset typically providing tighter information estimates.

\section{Results}

\subsection{Information estimation on simulated data}

We start by considering three simple chemical reaction networks for which we can obtain exact information values using the model-based approach outlined in  Methods Section \ref{sec:mexact}. This will allow us to precisely assess the performance of decoding-based model-free estimates, and systematically study the effects of time discretization, the number of sample trajectories, and the number  of distinct discrete inputs, $q$.

The three examples are all  instances of a simple molecular birth-death process, where molecules of $\tilde{X}$ are created and destroyed with rates $\alpha$ and $\beta$, respectively:
\begin{equation}
\xrightarrow{\alpha(U)} \tilde{X} \xrightarrow{\beta(U)} \varnothing .
\label{eq:Case2ReactionNetwork}
\end{equation}
The reaction rates, $\alpha$ and $\beta$, will depend in various ways on the input, $U$, and possibly time, as specified below. Given an initial condition, $x(t=0)$, the production and degradation reactions generate continuous-time stochastic trajectories, $x(t)$, recording the number of molecules of $\tilde{X}$ at every time $t\in[0,T]$, according to the Chemical Master Equation~(\ref{eq:CME}). These trajectories, or their discretized representations, are considered as the ``outputs'' of the example reaction networks, defining the mutual information $I(X;U)$ that we wish to compute. In all three examples we start with the simplest case, where the random variable $U$ can only take on two possible values, $u^{(1)}$ and $u^{(2)}$, with equal probability, $p_U(u^{(1)})=p_U(u^{(2)})=0.5$.

\begin{figure}[tb] 
	\centering
 	\includegraphics[width=1\textwidth]{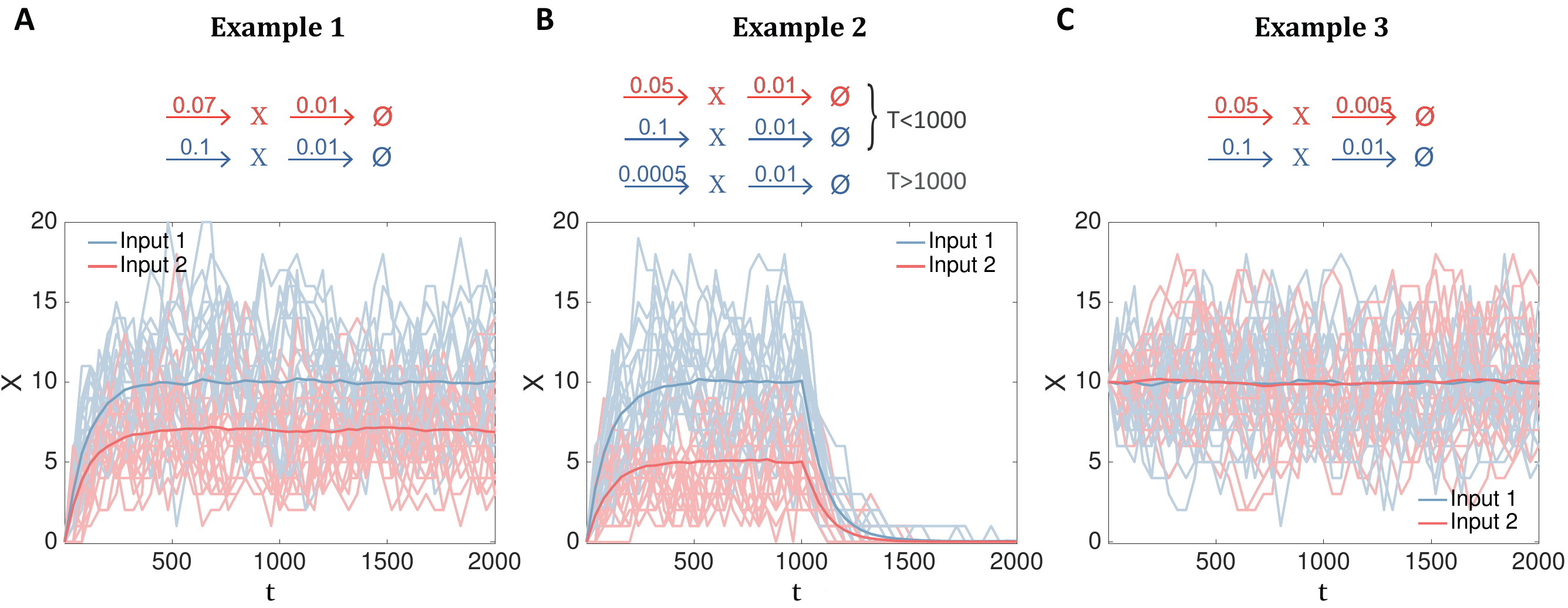}    
	\caption{{\bf Example biochemical reaction networks and their behavior.} Three example birth-death processes, specified by the reactions in the top row for each of the two possible inputs ($u^{(1)}$ in blue, $u^{(2)}$ in red), stylize simple behaviors of biochemical signaling networks. {\bf (A)} Input is encoded in both the transient approach to steady state and the steady state value. {\bf (B)} Input  is encoded in the magnitude of the transient response which is subsequently adapted away. {\bf (C)} Input is encoded only at the level of temporal correlations of the response trajectory. Bottom row shows example trajectories generated using the Stochastic Simulation Algorithm for the copy number of $\tilde{X}$ molecules, $t\in[0,2000]$, for each network and the two possible inputs (light blue, light red); while plotted as a connected line for clarity, each trajectory represents molecular counts and is thus a step-wise function taking on only integer or zero values. Dark blue, red lines show the conditional means over $N=1000$ trajectory realizations.}
	\label{fig:CasesPaths}
\end{figure}

\begin{itemize}
\item {\bf Example 1.} In this case, $x(t=0)=0$, $\beta=0.01$, independent of the input $U$, and the production rate depends on the input as $\alpha(u^{(1)})=0.1$, $\alpha(u^{(2)})=0.07$. Here, the steady state is given by Poisson distribution with mean number of molecules $\langle x(t\rightarrow \infty)\rangle = \alpha/\beta$. Steady-state is approached exponentially with the timescale that is the inverse of the degradation rate, $\beta^{-1}$. These dynamics stylize a class of frequently observed biochemical responses where the steady-state mean expression level encodes the relevant input value. Even if the stochastic trajectories for the two possible inputs are noisy as shown in Fig~\ref{fig:CasesPaths}A, we expect that the mutual information will climb quickly with the duration of the trajectory, $T$, since (especially in steady state) more samples provide direct evidence about the relevant input already at the level of the mean trajectories. 
\item {\bf Example 2.} In this case, $x(t=0)=0$, $\beta=0.01$, independent of the input $U$, and the production rate depends on the input as $\alpha(u^{(1)}, t)=0.1$, $\alpha(u^{(2)}, t)=0.05$ for all $t<1000$, while for $t\geq 1000$ the production rate is very small and independent of input, $\alpha(u,t)=5\cdot 10^{-4}$. In the early period, this network approaches input-dependent steady state with means whose differences are larger than in Example 1, but the difference decays away for $t>1000$ as the network settles towards vanishingly small activity for both inputs, as shown in Fig~\ref{fig:CasesPaths}B. These dynamics stylize a class of transient biochemical responses that are adapted away even if the input state persists. In this case, lengthening the observation window $T$ will not provide significant increases in information.
\item {\bf Example 3.} In this case, $x(t=0)=10$. All reaction rates depend on the input, $\alpha(u^{(1)})=0.1, \alpha(u^{(2)})=0.05,\beta(u^{(1)})=0.01,\beta(u^{(2)})=0.005$, and are chosen so that the mean $\langle x(t)\rangle=10$ is constant across time and equal for both conditions, as shown in Fig~\ref{fig:CasesPaths}C. In this difficult case, inputs cannot be decoded at the level of mean responses but require sensitivity to at least second-order statistics of the trajectories. Specifically, signatures of the input are present in the autocorrelation function for $x$: the timescale of fluctuations and mean-reversion is two-fold faster for $u^1$ than $u^2$. While this case is not frequently observed in biological systems, it represents a scenario where, by construction, no information about the input is present at the level of single concentration values and having access to the trajectories is essential. Because there is no difference in the mean response, we expect linear decoding methods to provide zero bits of information about the input. This case is also interesting because of the recent focus on pulsatile stationary-state dynamics in biochemical networks~\cite{Dalal2014}. These pulses, reported for transcription factors such as Msn2, NF-$\kappa$B, p$53$, etc., occur stochastically and, when averaged over a population of desynchronized cells, can yield a flat and featureless mean response. Information about the stimulus could, nevertheless, be encoded in either the frequency, amplitude, or other shape parameters of the pulses. While a generative description of such pulsatile dynamics goes beyond a birth-death process considered here, from the viewpoint of decoding, both pulsatile signaling and our example present an analogous problem, where the mean response is not informative about the applied input.
\end{itemize}

{\bf Exact information approximations and bounds for continuous and discrete trajectories.} 
Armed with the full stochastic model for the three example reaction networks, we can compute the mutual information, $I^*_{\rm exact}(X,U)$, between the continuous-time stochastic trajectories and the (binary) input variable $U$, following Eq~(\ref{eq:Itrue}). This result depends essentially on the length of the observed trajectory, $t\in[0,T]$, since $T$ controls the number of observed reaction events and thus the accumulation of evidence for one or the other alternative input. As the approximation is implemented by Monte-Carlo averaging of exact log probabilities for the response trajectories, its variance will depend on the number of sample trajectories generated by the SSA. Because these information values will represent the ``gold truth'' against which to evaluate subsequent estimators, we choose a large number of $N=1000$ trajectory realizations per input condition, and verify the tightness of the exact Monte Carlo approximation by computing the standard deviation over 20 independent re-runs of the approximation procedure. 

\begin{figure}[tb] 
	\centering	
	\includegraphics[width=\textwidth]{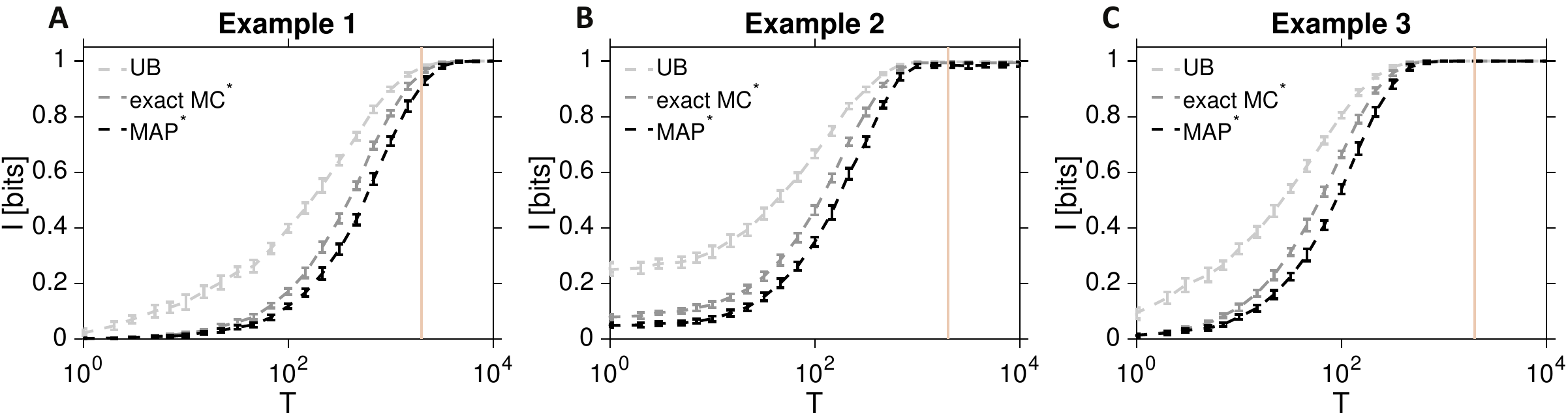}
	\caption{{\bf Information about inputs encoded by complete response trajectories of the example biochemical reaction networks.} Exact Monte Carlo approximation for the information, $I^*_{\rm exact}(X;U)$, is shown for Example 1 {\bf (A)}, Example 2 {\bf (B)}, and Example 3 {\bf (C)} from Fig~\ref{fig:CasesPaths} in dashed dark gray line; error bars are standard deviations across 20 replicate estimations, each computed over $N=1000$ independently generated sample trajectories per input condition. Information is plotted as a function of the trajectory duration, $T$; yellow vertical line indicates $T=2000$ as a representative duration used in further analyses below, at which most of the information about input is in principle available from the response trajectories of our systems.   $I_{\rm MAP}^*(\hat{U};U)$ (dashed black line) is the optimal decoding lower bound, and $I^*_{\rm UB}$ (dashed light gray) is the upper bound on the information, computed by applying Eq~(\ref{eq:UpperBound}).} 
	\label{fig:InfoCont}
\end{figure}

Figure~\ref{fig:InfoCont} shows how the exact Monte Carlo information computation depends on the trajectory duration, $T$, for each of the three example cases. As expected, the information increases monotonically with $T$ towards the theoretically maximal value of 1 bit, corresponding to perfect information about two \emph{a priori} equally likely input conditions. The exact shape of the information curve depends on the shape of the mean trajectory, as well as on its variance and higher-order statistics: for example, even though the two inputs for Example 1 are most distinct at the level of mean responses for later $T$ values, the noise is higher compared to Example 2, such that at $T=2000$ there is more total information in trajectories of Example 2 than Example 1. Conversely, even though the trajectories in Example 3 do not differ at the level of the mean at all, they still carry all information about the relevant input once sufficiently long trajectories can be observed  (and assuming full knowledge of the reaction network is available). 

One can similarly compute the Bayes-optimal or MAP decoding bound using Eq~(\ref{eq:mapdecoder}) for continuous trajectories. This quantifies the ultimate accuracy limit with which each single observed trajectory can be decoded into the input that gave rise to it. As demonstrated in Fig~\ref{fig:InfoCont} in dashed black line and consistent with the Data Processing Inequality requirements outlined in the Methods, $I_{\rm MAP}^*(\hat{U};U)\leq I^*_{\rm exact}(X;U)$. Equality is not reached because the optimal use of the channel requires block coding schemes, in contrast to our setting where different inputs are sequentially sent through the biochemical network and immediately decoded. The observed gap between the MAP optimal decoding estimate and the true information appears to be small in each of the three cases; one can upper-bound the gap itself by an improvement over the standard Feder-Merhav calculation following Methods Section~\ref{methods:fm}. While the resulting upper bound on information, $I_{\rm UB}$, is not tight in this case, it nevertheless provides a control of how far optimal decoding could be from the true information estimate, a question that has repeatedly worried the neuroscience community facing similar problems~\cite{Borst1999}. It is worth noting that if MAP decoder can tractably be computed, so can the upper bound, irrespective of the dimensionality of the space of responses, $X$. 

\begin{figure}[tb] 
	\centering
	\includegraphics[width=1\textwidth]{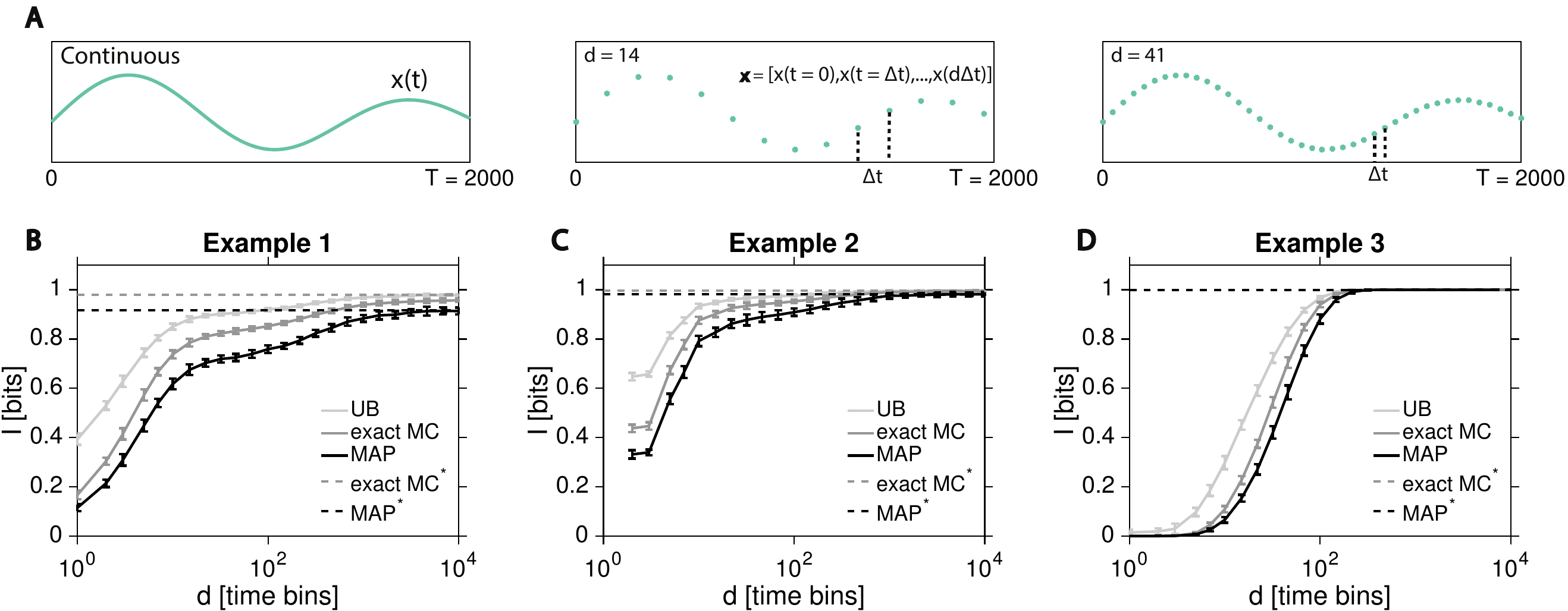}
	\caption{{\bf Information loss due to temporal sampling.} {\bf (A)} Schematic representation of the resampling of a continuous-time response trajectory (left) at $d=14$ (middle) or $d=41$ (right) equally spaced time points. Resampled response trajectories are represented as $d$-dimensional real vectors, $\mathbf{X}\in \mathbb{R}^d$, for the case of a single output chemical species. {\bf (B--D)} Exact Monte Carlo information approximations for discrete trajectories, $I_{\rm exact}(\mathbf{X};U)$ (dark solid gray), optimal decoding lower bound, $I_{\rm MAP}(\hat{U};U)$ (dark solid black), and the upper bound, $I_{\rm UB}$ (light solid gray) are plotted as a function of $d$. Continuous-time limits from Fig~\ref{fig:InfoCont} are shown as horizontal lines: $I_{\rm exact}^*(X;U)$ (dashed dark gray), $I_{\rm MAP}^*(\hat{U};U)$ (dashed black). Error bars as in  Fig~\ref{fig:InfoCont}.
	} 
	\label{fig:CasesInfoDisc}	
\end{figure}

Figure~\ref{fig:InfoCont} summarizes the absolute limits on information transmission and optimally decodable information, for each of our three example networks. These values are limits inasmuch as they assume that every reaction event can be observed and recorded with infinite temporal precision, and that the encoding stochastic process is perfectly known. While it is interesting to contemplate whether biological systems themselves could compute with or act on singular, precisely-timed reaction events and thus make optimal use of the resulting channel capacity (mimicking the debate between spike timing code and spike rate code in neuroscience), our primary focus here is to estimate information flows from experimental data. Typically, experiments record the state of the system---e.g., concentration of signaling molecules---in discretely sampled time. To explore the effects of time discretization, we first fix the observation length for our trajectories to $T=2000$, sufficiently long that the trajectories in principle contain more than $90\%$ of the theoretically maximal information for each of the three example cases. We then resample the trajectories on a grid of $d$ equally spaced time points, as illustrated in Fig~\ref{fig:CasesInfoDisc}A.

Figures~\ref{fig:CasesInfoDisc}B-D compare the exact Monte Carlo information approximation for discrete trajectories, $I_{\rm exact}(\mathbf{X};U)$, MAP lower bound for discrete trajectories, $I_{\rm MAP}(\hat{U};U)$, and the corresponding upper bound, $I_{\rm UB}$, to the theoretical limits from Fig~\ref{fig:InfoCont} obtained using continuous trajectories. In line with the chain of inequalities in Eq~(\ref{eq:finaldpi}), information in discretely resampled trajectories is lower than the true information in continuous trajectories, but converges to the true value as $d\rightarrow \infty$. In particular, once the discretization timestep $T/d$ is much lower than the inverse of the fastest reaction rate in the system, discretization should incur no significant loss of information. In practice, however, high sampling rate (large $d$) limit has significant drawbacks: first, it is technically difficult to take snapshots of the system at such high rates (e.g., due to fluorophore bleaching); second, the fast dynamics of the reaction network may be low-pass filtered by the readout process (e.g., due to fluorophore maturation time, or slower downstream reaction kinetics); and third, for model-free approaches high $d$ implies that decoders need to be learned over input spaces of high dimensionality, which could be infeasible given a limited number of experimentally recorded response trajectories. In previous work~\cite{Hansen2015,Hafner2017}, trajectories were typically represented as $d\approx 1\sim 100$ dimensional vectors, which in our examples would capture $\sim 80\%$ or more of the theoretically available information. It is likely that this can be improved further with smart positioning of the sampling points and that not all theoretically available information could actually be accessed by the organism itself, suggesting that typically used discretization approaches have the potential to capture most of the relevant information in the responses. What is important for the analysis at hand is that given the dimensionality $d$ of the discretized response trajectories, MAP decoder  is guaranteed to reach the minimal decoding error among all possible decoders, and will turn out to be a relevant benchmark, by yielding the highest information, $I_{\rm MAP}(\hat{U};U)$, in Figs~\ref{fig:CasesInfoDisc}B-D among all decoders considered. In what follows, we will examine how various model-free decoding estimators approach this limit, as a function of $d$ and the number of sample trajectories, $N$.

\begin{figure}[tb] 
	\centering	
	\includegraphics[width=\textwidth]{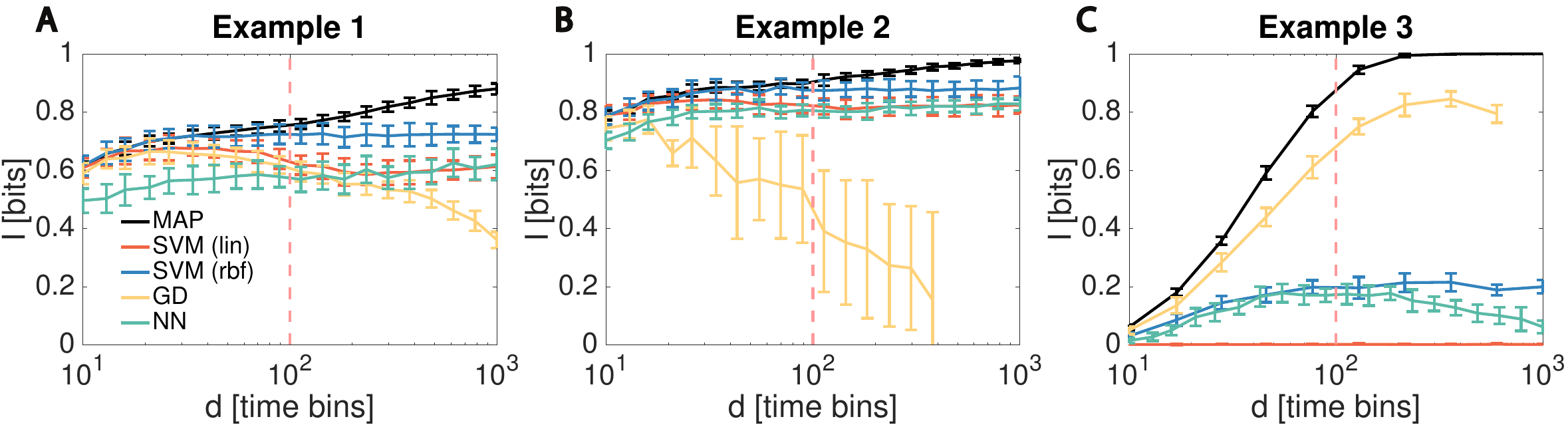}
	\caption{{\bf Performance of decoding-based estimators depends on the dimensionality of the response trajectories.} Performance of various model-free decoding estimators (colored lines) for   Examples 1 {\bf (A)}, 2 {\bf (B)}, 3 {\bf (C)}, respectively, compared to the MAP bound, $I_{\rm MAP}$ (black line), as a function of input trajectory dimension, $d$. In all cases, the number of sample trajectories per input condition is $N=1000$, error bars are std over $20$ replicate estimations. Decoding estimators: linear SVM, $I_{\rm SVM (lin)}$ (orange); radial basis functions SVM, $I_{\rm SVM (rbf)}$ (blue); the Gaussian decoder with diagonal regularization (see Fig~\ref{fig:CS3GDReg}), $I_{\rm GD}$ (yellow); multi-layer perceptron neural network, $I_{\rm NN}$ (green). Dashed vertical orange line marks the $d\leq 100$ regime typical of current experiments. } 
	\label{fig:CasesInfoDecodeNtb}	
\end{figure}

\begin{figure}[tb] 
	\centering
	\includegraphics[width=\textwidth]{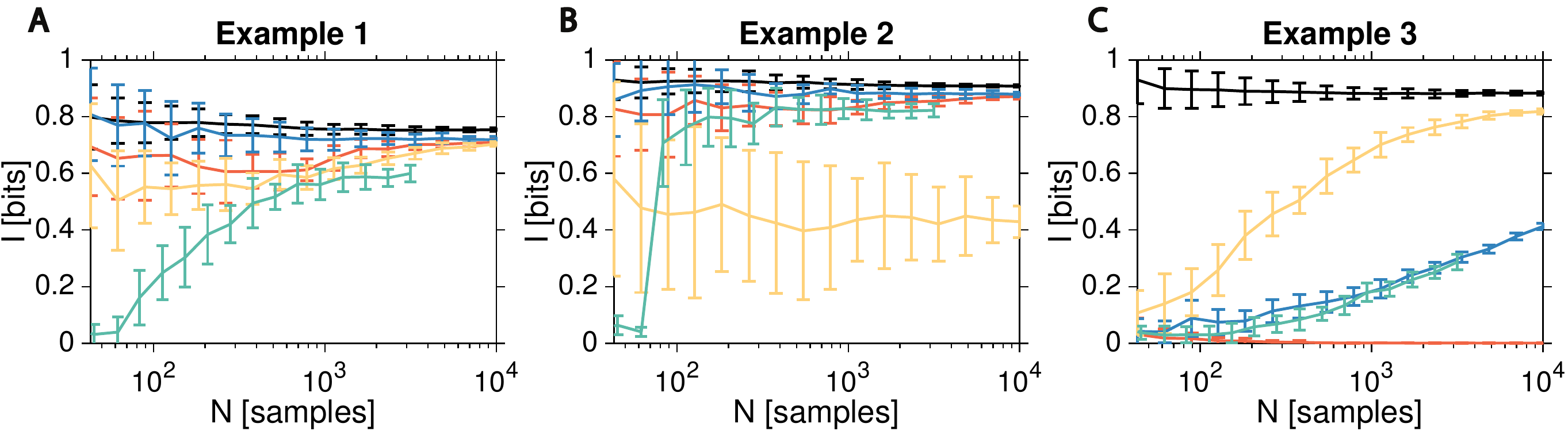}
	\caption{{\bf Convergence of decoding-based estimator performance with increasing number of response trajectory samples.}  Performance of various model-free decoding estimators (colored lines) for Examples 1 {\bf (A)}, 2 {\bf (B)}, 3 {\bf (C)}, respectively, compared to the MAP bound, $I_{\rm MAP}$ (black line), as a function of the number of samples, $N$, per input condition. Response trajectories are represented as $d=100$ dimensional vectors. Plotting conventions as in Fig~\ref{fig:CasesInfoDecodeNtb}. } 
	\label{fig:CasesInfoDecodeN}	
\end{figure}

{\bf Performance of decoding-based estimators.}
After establishing our model-based ``gold standard'' for decoding-based estimators acting on trajectories represented in discretized time,  $I_{\rm MAP}(\hat{U};U)$, we turn our attention to the performance comparison between various model-free algorithms. The results are summarized in Fig~\ref{fig:CasesInfoDecodeNtb}, which shows how  estimator accuracy depends on the dimensionality of the problem, $d$, given a fixed number, $N=1000$, of  sample trajectories per input condition. In contrast, Fig~\ref{fig:CasesInfoDecodeN} assumes a fixed dimensionality $d$ of trajectory vectors and explores how the estimator performance depends on the number of samples, $N$.

Figures~\ref{fig:CasesInfoDecodeNtb} and \ref{fig:CasesInfoDecodeN} lead us to the following conclusions:
\begin{itemize}
\item {\bf Nonlinear SVM} using the radial basis functions (rbf) kernel performs best for Examples 1 and 2. Regardless of the number of samples, $N$, or the number of time bins, $d$, its estimates are very close to $I_{\rm MAP}$, especially for the relevant regime $d\sim 10-100$. Even for higher $d$, the estimator shows hardly any overfitting and thus stable performance, a feature we have observed commonly in our  numerical explorations. The estimator is sample efficient, typically providing estimates with smallest error bars. 
\item {\bf Linear SVM} slightly underperforms kernelized SVM on Examples 1 and 2, and---as expected---completely fails on the linearly inseparable Example 3. Interestingly, even though more expressive, kernelized SVM seems to incur no generalization cost relative to linear SVM even at low number of samples. For all examples we tested, kernelized SVM thus appears to be a method of choice; linear SVM, however, is still useful as a benchmark to measure what fraction of the information is linearly decodable from the signal.
\item {\bf The Gaussian decoder} has the best performance on Example 3, is competitive for low $d$ for Example 1, and doesn't perform satisfactory for Example 2. As shown in Fig~\ref{fig:CS3GDReg}, regularized estimation of covariance matrix appears crucial for good performance, but smoothing of the originally discrete trajectory does not help. Even with regularization, this estimator is not sample efficient for Example 1: the trajectories are linearly separable without a full estimate of the covariance (as evidenced by the success of the linear SVM), yet the Gaussian decoder requires one to two orders of magnitude more samples to match the linear decoder performance. This drawback turns into a benefit for Example 3: the Gaussian assumption can be viewed as a prior that second-order statistics are important for decoding (which is correct in this case). Kernelized SVM and the neural network, while more general, need to learn from many more training samples to zero in on these features, and fail to reach the Gaussian decoder performance even for $N=10^4$. We hypothesized that the failure of the Gaussian decoder on Example 2 is due to the difficulty of the Gaussian approximation to capture the period $T>1000$ when the mean number of $\tilde{X}$ is close to zero: here, first, the Gaussian assumption must be strongly violated, and, second, the estimation of (co)variance from finite number of samples is close to singular due to the small number of reaction events in this period. Even though the $T>1000$ epoch is not informative about the input, a badly conditioned decoder for this epoch can actually adversely affect performance. We confirmed this hypothesis by building the Gaussian decoder restricted to $T<1000$ that reliably extracted $\geq 0.8$ bits of information in Example 2, close to the MAP decoding bound and the performance of SVM-based estimators.
\item {\bf Neural network decoder} reaches a comparable performance on Examples 2 and 3 to the SVMs, but fails to be competitive for the simple Example 1. This is most likely because this estimator is sample inefficient, as implied by its continual increase in performance with $N$ that did not saturate at highest $N$ we tried.  Given their expressive power, neural network decoders should be viewed as the opposite benchmark to the linear decoders: they have the ability to pick up complex statistical structures but only with a sufficient number of samples. Indeed, as we will see subsequently for applications to real data, neural networks can match and exceed the performance of SVMs. We emphasize that we used a neural network with a fixed architecture for all three examples on purpose, to make results comparable across examples; the performance can likely be improved by optimizing the architecture separately for each estimation problem.
\end{itemize}

\begin{figure}[tb] 
	\centering
	\includegraphics[width=.8\textwidth]{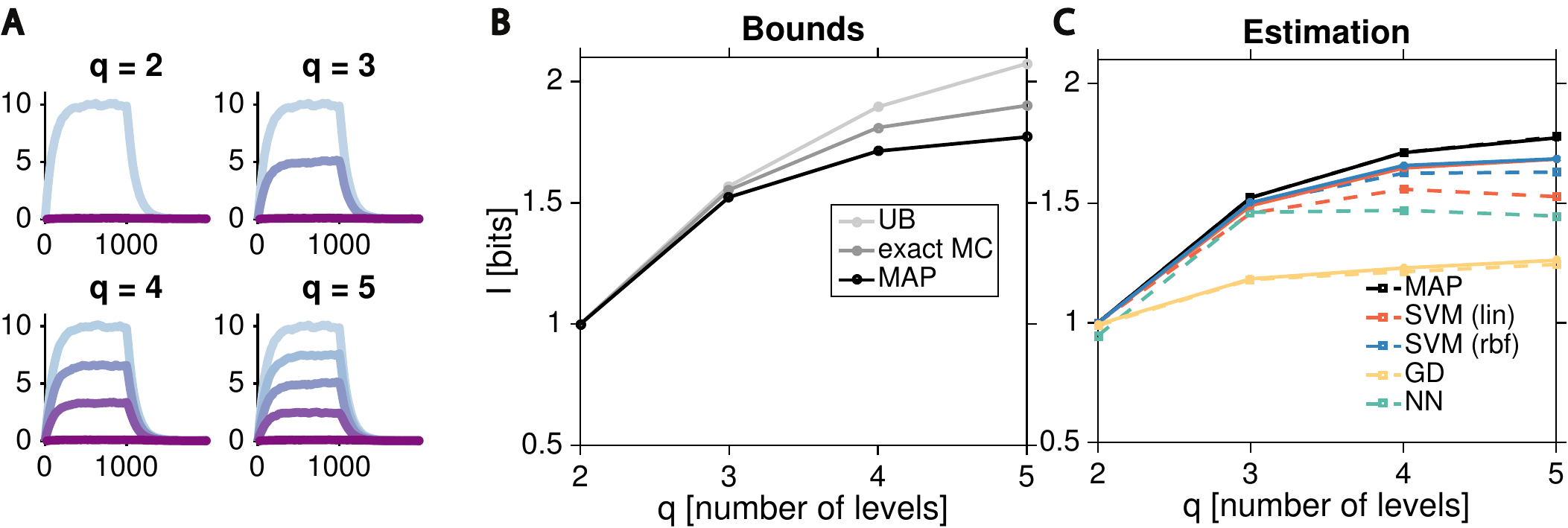}	
	\caption{{\bf Information estimation for multilevel inputs.} {\bf (A)} Extension of Example 2 from Fig~\ref{fig:CasesPaths}B to $q=2,\dots,5$ discrete inputs. We chose the inputs such that the response for the system at $T<1000$ converges towards $q$ equally spaced levels with the same dynamic range as the original example; dynamics at $T\geq 1000$ remain unchanged from the original Example 2. {\bf (B)} Model-based information bounds as a function of the number of input levels for trajectories represented as $d=100$ dimensional vectors: exact Monte Carlo calculation (dark gray), MAP decoding bound (black), upper bound (light gray). {\bf (C)} Performance of model-free estimators, as indicated in the panel, compared to the MAP bound (black). Dashed lines show estimations using $N=10^3$ sample trajectories per condition, solid lines using $N=10^4$ samples per condition; in both cases, we show an average over $20$ independent replicates, error bars are suppressed for readability.} 
	\label{fig:LevelsInfo}	
\end{figure}

{\bf Multilevel information estimation.} We next asked whether our conclusions hold also when the space of possible inputs is expanded beyond binary, assuming that $U$ can take on $q$ distinct values with equal probability, i.e., $p_U(u)=1/q$. We focused on Example 2, and constructed cases for $q=2,\dots,5$ such that the production rate $\alpha$ for $0<T<1000$ takes on $q$ uniformly spaced values between $0$ and the maximal rate equal to $\alpha=0.1$ used in Fig~\ref{fig:CasesPaths}B. In effect, this ``tiles'' the original, two-state-input dynamic range uniformly with $q$ input states, as illustrated in Fig~\ref{fig:LevelsInfo}A.

Our expectation is that with increasing $q$, the information should increase, but slowly saturate as reliable distinctions between nearby input levels can no longer be made due to the intrinsic biochemical stochasticity. This is indeed what we see in Fig~\ref{fig:LevelsInfo}B, which shows the exact information, the MAP lower  bound and the upper bound. Consistent with our findings for two-input case, SVM using radial basis functions remains the estimator of choice for all values of $q$, followed by the linear SVM and then the neural network decoder, as shown in Fig~\ref{fig:LevelsInfo}C.

{\bf Performance comparisons with model-free information approximations.} There exist many algorithms for estimating information directly, without making use of the decoding lower bound. The best known estimator for continuous signals is perhaps the k-nearest-neighbor (knn) estimator~\cite{Kraskov2004}. We have also introduced estimators based on parametric assumptions about the response distribution, such as the Gaussian approximation (Methods Section~\ref{sec:mfe}); both belong in the family of binless approximations, which act directly on real-valued response vectors. In contrast, binning approximations first discretize the responses $\mathbf{X}$. The simplest such approach is perhaps the direct estimator of information or entropy~\cite{Strong1998}, and a good review is provided in Ref~\cite{Paninski2003}. We evaluated the performance of the Gaussian approximation to find that it can systematically overshoot the true information with a bias that is difficult to assess (Fig~\ref{fig:CS3GaussFiltering}); this appears to happen also in the regime where the biochemical noise should be small (relative to the mean), and the stochastic dynamics should be describable in terms of Langevin approximations with the resulting Gaussian response distributions. These approximations converge to the true solution in terms of their first and second moments, yet do not seem to lead to unbiased estimate for the entropies and thus the mutual information. In contrast to the Gaussian decoder, Gaussian approximation should not be used without a better understanding of its bias and applicability. 

\begin{figure}[tb] 
	\centering
	\includegraphics[width=.5\textwidth]{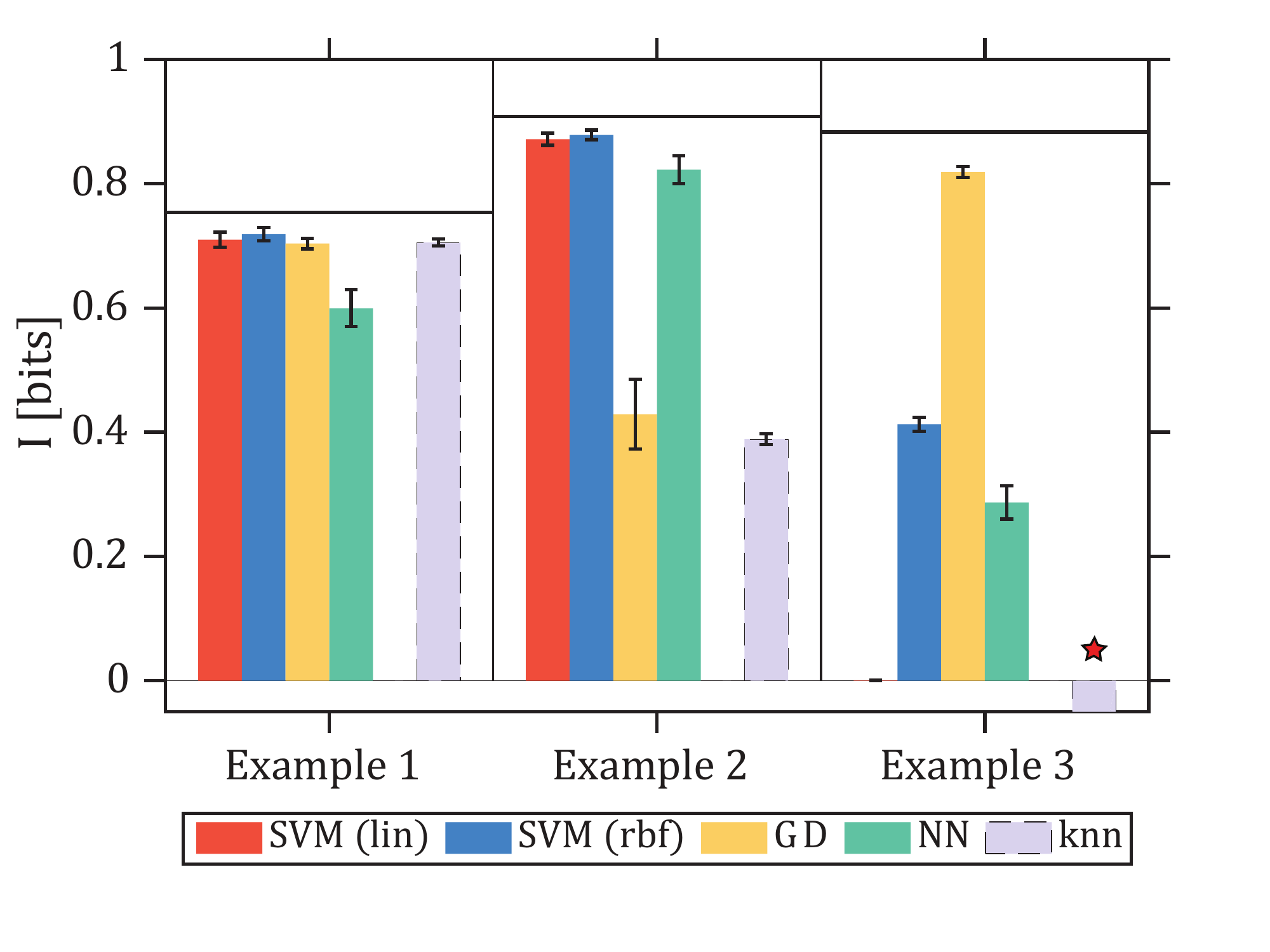}	
	\caption{{\bf Comparison of decoding-based and knn information estimators.} Information estimates for decoding-based (color bars) and knn (gray bar) algorithms (here we set $k=1$, for further details of knn estimation, see SI Fig~\ref{fig:knn_cs123}). Note that knn is not a decoding estimator and thus could exceed $I_{\rm MAP}(\hat{U};U)$ (shown as a horizontal black line for each of the three example cases) to approach the exact $I_{\rm exact}(\mathbf{X};U)$.  Here we use trajectories discretized over $d=100$ time bins, and  $N=10^4$ trajectory samples per input. The performance of knn can be substantially improved by adding a small amount of gaussian noise to the trajectory samples; its resulting performance as a function of $N$ and $d$ is shown in Fig~\ref{fig:knn_cs123}. Red star denotes the failure of knn on Example 3 where substantially negative information values are returned (exact value not plotted).}
	\label{fig:BarplotCases}	
\end{figure}

We therefore decided to focus on the comparison of decoding estimators with knn, which has been used previously on data from biochemical signaling networks~\cite{Selimkhanov2014}. The results are shown in Fig~\ref{fig:BarplotCases}. K-nearest-neighbors performs well on the easy Example 1, and suffers drastic performance drop for Example 2, while crashing catastrophically by reporting negative values in Example 3. We reasoned that part of the difficulty may be the fact that synthetic trajectories for our Examples are defined over non-negative whole numbers only, whereas the knn assumes real valued vectors. This is confirmed by Fig~\ref{fig:knn_cs123} which shows that the knn performance can be substantially improved by adding a small amount of gaussian iid noise to every component of the response trajectory vectors, $\mathbf{X}$. This restores the knn performance in Example 2 close to that of the SVM-based estimators, but still produces close-to-zero bits of information for Example 3.

\subsection{Applications to real data}
To illustrate the use of our estimators in a realistic context, we analyzed data from two previously published papers. The first paper focused on the representation of environmental stress in the nuclear localization dynamics of several transcription factors (here we focus on data for Msn2, Dot6, and Sfp1) in budding yeast~\cite{Granados2018}. The second paper studied information transmission in biochemical signaling networks in mammalian cells (here we focus on data for ERK and Ca$^+$)~\cite{Selimkhanov2014}. In both cases, single-cell trajectory data were collected in hundreds or thousands of single cells sampled at sufficient resolution to represent the trajectories discretized at tens to hundreds of timepoints. Similarly, both papers estimate the information transmission in trajectories about a discrete number of environmental conditions: Ref~\cite{Granados2018} uses the linear SVM approach presented here, while Ref~\cite{Selimkhanov2014} uses the knn estimator. This makes the two datasets perfectly suited for estimator comparisons. We further note that in both datasets the trajectories can be divided into two response periods: the early ``transient'' response period when the external condition changes, and the late ``near steady-state'' response period. Typically, the transient dynamics exhibit clear differences in the trajectory means between various conditions, reminiscent of our Example 1 or early Example 2; in contrast, in the late period the response may have been adapted away, or the stimulus could be encoded only in higher-order statistics of the traces, reminiscent of the late period in Example 2 or Example 3. 

\begin{figure}[tb]
	\centering
	\includegraphics[width=.7\textwidth]{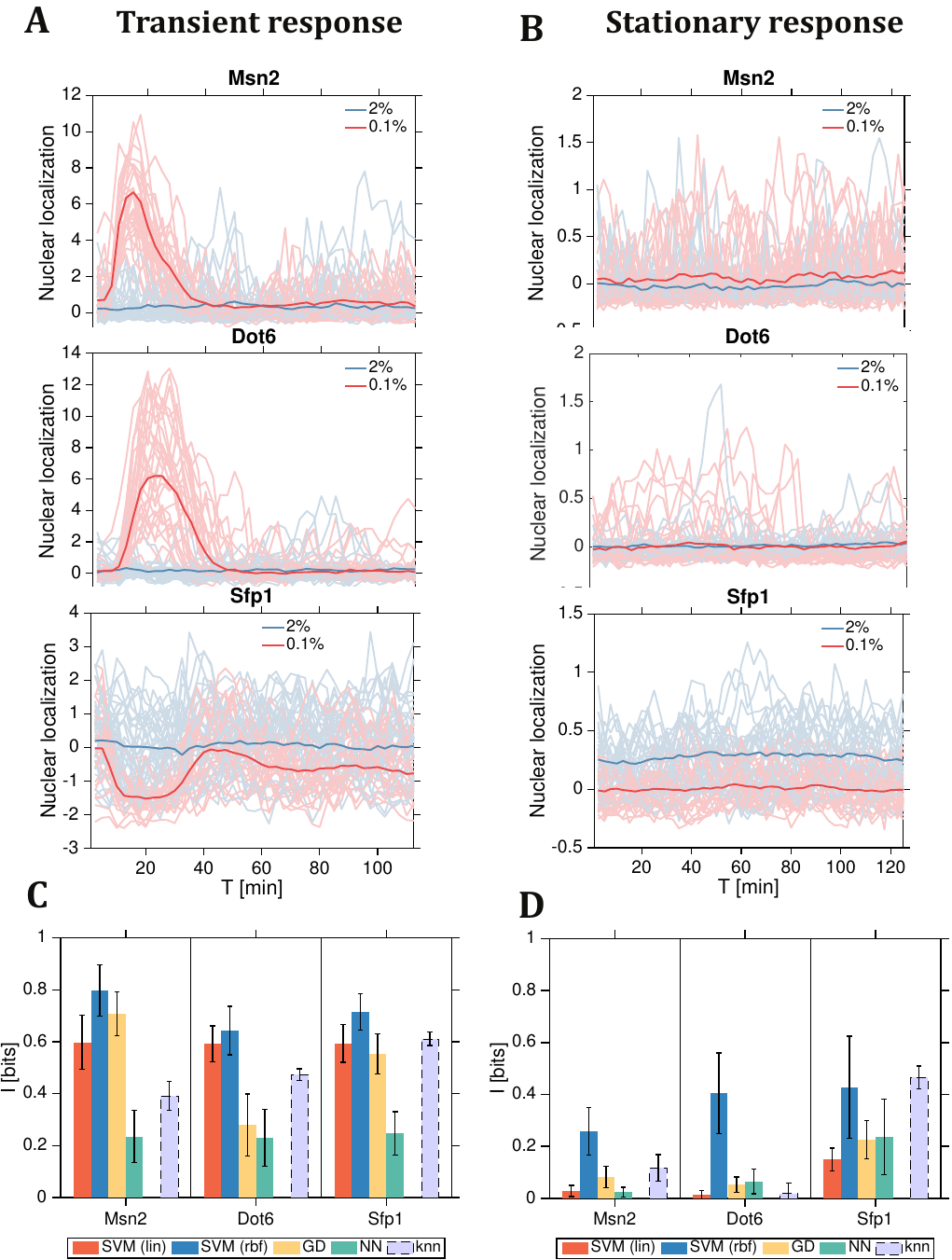}
	\caption{{\bf Two-level mutual information estimates from single-cell time-series data for nuclear translocation of yeast transcription factors.} {\bf (A, B)} Data replotted from Ref~\cite{Granados2018} for Msn2 (top row), Dot6 (middle row), and Sfp1 (bottom row); early transient responses (A) after nutrient shift at $t=0$ min from glucose rich ($2\%$, blue traces) to glucose poor ($0.1\%$, red traces) medium are shown in the left column, stationary responses (B) are collected after cells are fully adapted to the new medium. Sampling frequency is $2.5$ min, $d=45$, and the number of sample trajectories per nutrient condition is $N=100$. Thin lines are individual single cell traces, solid lines are population averages. {\bf (C, D)} Information estimates for the transient (left, C) and stationary (right, D) response periods. Colored bars use model-free decoding-based estimators as indicated in the legend, gray bar is the knn estimate; error bars computed from estimation bootstraps by randomly splitting the data into testing and training sets.
 }
	\label{fig:InfoEstData}	
\end{figure}

\begin{figure}[tb]
	\centering
	\includegraphics[width=.7\textwidth]{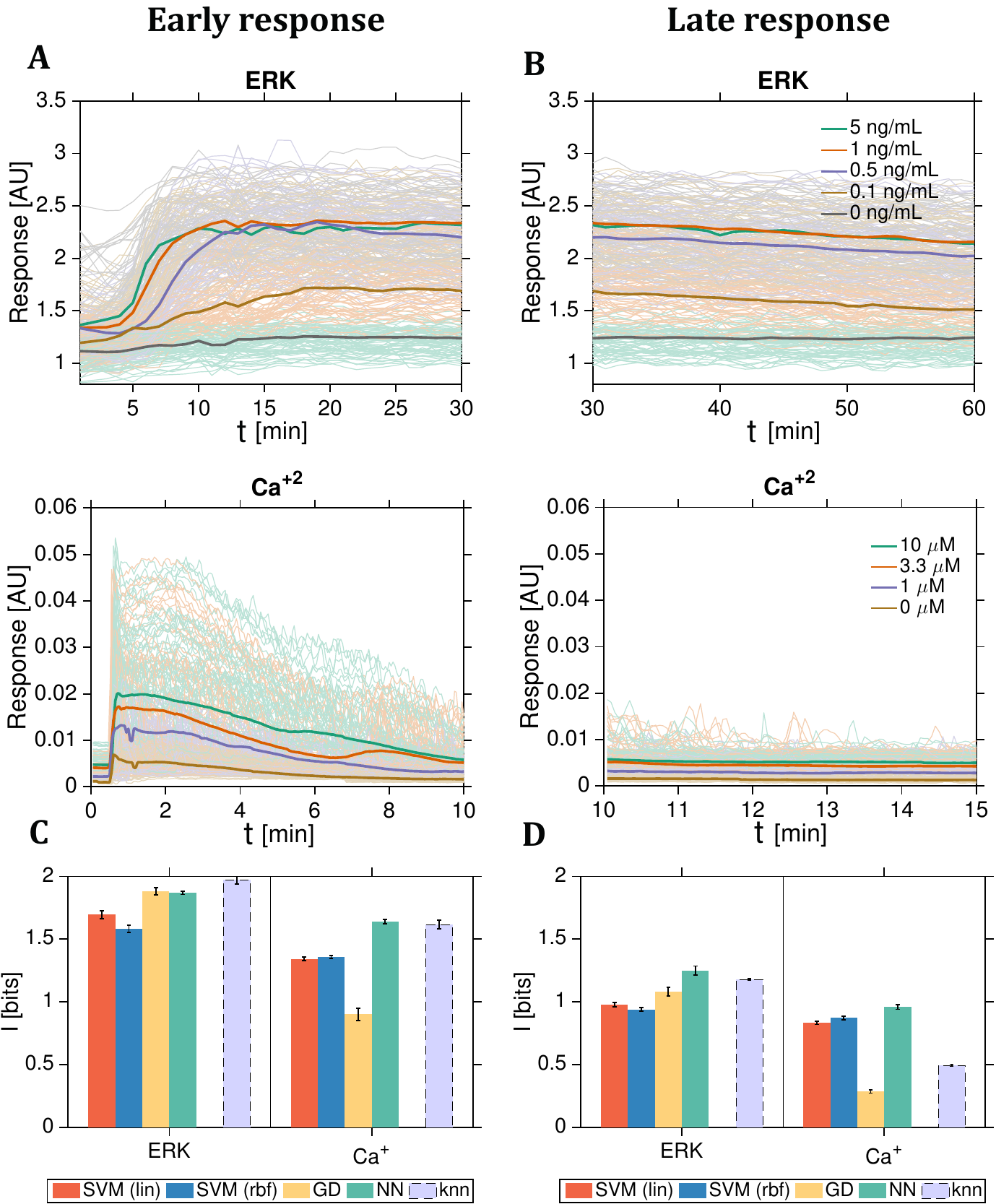}
	\caption{{\bf Multilevel mutual information estimates from single-cell time-series data for mammalian intracellular signaling.} Data replotted from Ref~\cite{Selimkhanov2014} for ERK (top row) and Ca$^+$ (bottom row). {\bf (A)} Early transient responses  after addition of 5 different levels of EGF for ERK (or 4 different levels of ATP for Ca$^+$, respectively) at $t=0$ min, as indicated in the legend. {\bf (B)} In the  late response most, but not all, of the transients have decayed. Data for ERK: $N=1678$ per condition, $T=30$ min ($d=30$) for early response and $T=30$ min ($d=30$) for late response. Data for Ca$^+$: $N=2995$ per condition, $T=10$ min ($d=200$) for early response and $T=5$ min ($d=100$) for late response.  Plotting conventions as in Fig~\ref{fig:InfoEstData}. }
	\label{fig:InfoEstDataERK_CA}	
\end{figure}

Figure~\ref{fig:InfoEstData} shows the raw data and summarizes our estimation results for the early and late response periods for the three translocating factors in yeast that report on the change from $2\%$ glucose rich medium to $0.1\%$ glucose poor stress medium. Figure~\ref{fig:InfoEstDataERK_CA} similarly shows the raw data and estimation results for the early and late response periods for the signaling molecules in mammalian cells responding to multilevel inputs.  

Consistent with the published report~\cite{Granados2018}, transient response in yeast nuclear localization signal can be decoded well with the linear SVM estimator that yields about 0.6 bits of information per gene about the external condition. Kernelized SVM outperforms the linear method slightly by extracting an extra 0.1-0.2 bits of information, while knn underperforms the linear method significantly for Msn2 and Dot6 (but not for Sfp1). The Gaussian decoder estimate shows a mixed performance and the neural network estimate is the worst performer, most likely because the number of  samples here is only $N=100$ per input condition and neural network training is significantly impacted. 

It is interesting to look at the stationary responses in yeast which haven't previously been analyzed in detail. First, low estimates provided by linear SVM for Msn2 and Dot6 imply that information in the stationary regime, if present, cannot be extracted by the linear classifier. Second, the Gaussian decoder also performs poorly in the stationary regime, potentially indicating that the relevant features are encoded in higher-than-pairwise order statistics of the response (e.g., pulses could be ``sparse'' features as in sparse coding~\cite{Olshausen2004}); it is, however, hard to exclude small number of training samples as the explanation for the poor performance of the Gaussian decoder. Third, K-nearest-neighbor estimator also yields low estimates, either due to small sample number or low signal-to-noise ratio, the regime for which knn method has been observed to show reduced performance~\cite{Khan2007}. A particularly worrying feature of the knn estimates is their non-robust dependence on the length of the trajectory $T$. As Fig~\ref{fig:knn2} shows, the performance of knn peaks at $T\approx 50$ min and then drops, even well into unrealistic negative estimates for  $T\approx 400$ min (corresponding to the highest dimensionality $d=170$ of discrete trajectories). While it is possible to make an \emph{ad hoc} choice to always select trajectory duration at which the estimate peaks, the performance of kernelized SVM is, in comparison, extremely well behaved and increases monotonically with $T$, as theoretically expected. Finally, nonlinear SVM estimator extracts up to 0.4 bits of information about condition per gene, more than half of the information in the early transient period. This is even though on average the response trajectories for the two conditions, $2\%$ glucose and $0.1\%$ glucose, for Msn2 and Dot6 are nearly identical. For Sfp1 there is a notable difference in the mean response, which the linear estimator can use to provide a $\sim 0.15$ bits of information, yet still significantly below $\sim 0.4$ bits extracted by the nonlinear SVM. For both transient and stationary responses in yeast, our results are qualitatively in line with the expectations from the synthetic example cases---given the small number of trajectories, tightest and most robust estimates are provided by the decoding information estimator based on nonlinear (kernelized) SVM. Regardless of the decoding methodology and even without small sample corrections at $N=100$ trajectories per input, our estimates are not significantly impacted by the well-known information estimation biases thanks to the dimensionality reduction that decoding provides by mapping high dimensional trajectories $\mathbf{X}$ back into the space for inputs $U$ which is low dimensional; this is verified in Fig~\ref{fig:RandLabeling} by estimating the (zero) information in trajectories whose input labels have been randomly assigned.

Random pulses that encode stationary environmental signals have been observed for at least 10 transcription factors in yeast~\cite{Dalal2014} and for tens of transcription factors in mammalian cells~ \cite{Levine2013}. Recent studies investigated the role of the pulsatile dynamics in cellular decision-making~\cite{Albeck2013,Hafner2017}. Nevertheless, methods for quantifying the information encoded in stochastic pulses are still in their infancy. Our nonlinear SVM decoding estimates convincingly show that there is information to be learned at the single cell level from the stationary stochastic pulsing. An interesting direction for future work is to ask whether hand-crafted features of the response trajectories (pulse frequency, amplitude, shape, etc) can extract as much information from the trajectories as the generic SVM classifier: for that, one would construct for each response trajectory a ``feature vector'' by hand, compute the linear SVM decoding bound information estimate from the feature vectors, and compare that to the kernelized SVM estimate over the original trajectories. This approach is a generic and operationally-defined path for finding ``sufficient statistics'' of the response trajectories---or a compression of the original signal to the relevant set of features---in the information-theoretic sense.

A different picture emerges from the mammalian signaling network data shown in Fig~\ref{fig:InfoEstDataERK_CA}. The key difference here is the order of magnitude larger number of sample trajectories per condition compared to yeast data. Most of the information seems linearly separable in both the early and late response periods, as evidenced by the success of the linear SVM based estimator whose performance is not improved upon by the kernelized SVM (indeed, for early ERK response period linear SVM gives a slightly higher estimate than the nonlinear version). The big winner on this dataset is the neural-network-based estimator that yields the best performance in all conditions among the decoding-based estimators, likely owning to sufficient training data. As before, the Gaussian decoder shows mixed performance which can get competitive with the best estimators under some conditions. Lastly, knn appears to do well except on the late Ca$^+$ data (perhaps due to low signal-to-noise ratio). It also shows counter-intuitive non-monotonic behavior with trajectory duration $T$ in Fig~\ref{fig:ERK_CA}. Once again it is worth keeping in mind that knn is estimating the full mutual information which could be higher than the information decodable from single responses. 

\clearpage

\section{Discussion}
Increasing availability of single-cell time-resolved data should allow us to address open questions regarding the amount of information about the external world that is available in the time-varying concentrations, activation or localization patterns, and modification state of various biochemical molecules. Do full response trajectories provide more information than single temporal snapshots, as early studies suggest? Is this information gain purely due to noise averaging enabled by observing multiple snapshots, or---more interestingly---due to the ability of these intrinsically high-dimensional signals to provide a richer representation of the cellular environment? Can we isolate biologically relevant features of the response trajectories, e.g., amplitude, frequency, pulse shape, relative phase or timing, without {\em a priori} assuming what these features are? How can cells read out the environmental state from these response trajectories and how close to the information-theoretic bounds is this readout process?

Here, we made steps towards answering these questions by focusing on two related problems: first, if we are given a full stochastic description of a biochemical reaction network, under what conditions can we theoretically compute information transmission through this network and various related bounds; second, if we are given real data with no description of the network, what are tractable schemes to  estimate the information transmission. We show that when the complete state of the reaction network is observed and the inputs are discrete sets of reaction rates, there exist tractable Monte Carlo approximation schemes for the information transmission. These exact results that we compute for three simple biological network examples then serve to benchmark a family of decoding-based model-free estimators and compare their performance to the commonly-used knn estimator. We show that decoding-based estimators can closely approach the optimal decoder performance and in many cases perform better than knn, especially with typical problem dimensions ($d\sim 1-100$) and typical number of sample trajectories ($N\sim 10^2-10^3$). This is especially true when we ask about the combinatorial representation of the environmental state in the time trajectories of several jointly observed chemical species, as in our previous work~\cite{Granados2018}, where alternative information estimation methods usually completely fail due to the high dimensionality of the input space. 

It is necessary to emphasize the flexibility of the decoding approach: decoding-based information estimation is based directly on the statistical problems of classification (for discrete input variable, $U$) or regression (for continuous input variable, $U$), so any classification / regression algorithm with good performance can provide the basis for information estimation. Concretely, for problems in the low data regime (small $N$), linear or kernelized SVM approaches appear powerful, while at larger $N$ neural-network-based schemes can provide a better performance and thus typically a tighter information lower bound. In contrast to information approximations for which it is often impossible to assess their precision or bias (or even its sign) when the dimension, $d$, of the problem is large, the decoding approach yields a conservative estimate of the true information. Statistical algorithms underlying decoding-based estimations have the extra advantage that, (i), we may be able to gain biological insight by inspecting which features of the response carry the relevant stimulus information (e.g., by looking at the linear kernels or features that neural networks extract in their various layers); (ii), pick a  decoding algorithm based on features  previously reported as relevant (e.g., the Gaussian decoder for second-order statistics as in Example 3); (iii), estimate the information as a function of trajectory duration; and (iv), gain confidence in our estimates by testing their performance on withheld data. 

By construction, decoding-based estimators only provide a lower bound to the true information. This, however, could turn out to be a smaller problem in practice than it appears in theory, especially for biochemical reaction networks. First, our extension to the Feder-Merhav bound provides us with an estimate of how large the gap between the true information and the decoded estimation can be. The bound is not tight on our examples, and can only be applied when the optimal MAP decoder can be constructed~\cite{Tkacik2015}. Second, and perhaps more importantly, information that can be decoded after single input presentations is the quantity that is likely more biologically relevant than the true channel capacity, if the organisms are under constraint to respond to the environmental changes quickly. Typically, organisms across the complexity scale operate under speed-accuracy tradeoffs~\cite{heitz2014speed}: faster decisions based on noisy information lead to more errors and, conversely, with enough time to integrate sensory information errors can be reduced. When speed is at a premium or relevant inputs are sparse, decisions need to be taken after single input presentations.  In this case, decoding-based estimation should not be viewed as an approximate but rather as the correct methodology for the biological problem at hand. Of course, there is still the question of whether the model-free decoders that we use on real data can achieve a performance that is close to the optimal MAP decoder that represents the absolute performance limit. While there is no general way to answer this question, it appears that simple SVM decoding schemes work well when the response trajectories differ in their conditional mean, and neural networks as general approximators can be used to check for more complicated encoding features when data is plentiful. Unlike in neuroscience, there is much less clarity about what kind of read-out or decoding operations biochemical networks can mechanistically realize to mimic the functioning of our \emph{in silico} decoders, and it may be challenging to biochemically implement even arbitrary linear classification of response trajectories. Until experimentally shown otherwise, it thus appears reasonable to proceed with the assumption that environmental signals can be read out from the time-dependent internal chemical state with a simple repertoire of computations.

We conclude by emphasizing a simple yet important point. The decoding-based approach that we introduced here should also motivate us to look beyond methodological problems of significance and estimation, to truly biological problems of cellular decision making. Currently, data on biological regulatory processes is often analyzed by looking for ``statistically significant differences'' in the network response for, say, two possible network inputs. For example, one may report that the steady-state mean expression level of a certain gene is significantly larger in the stimulated vs unstimulated condition, with the statistical significance of the mean difference established through an appropriate statistical test that takes into account the number of collected population samples. While statistical significance is a necessary condition to validly report \emph{any} difference in the response, it is very different from the question of whether \emph{a single cell} could discriminate the two conditions given access only to its own expression levels. In caricature, population-level statistics tell us with what confidence we, as scientists having access to $N$ samples, can discriminate between conditions given some biological readout; decoding based information estimates, on the other hand, are relevant to the $N=1$ case of individual cells. We hope that further work along the latter path can clarify and quantify better the difficult constraints and conditions under which real cells need to act based on individual noisy readouts of their stochastic biochemistry.

\section{Acknowledgements}
We thank Alejandro Granados, Mihal Hledik, Julian Pietsch, and Christoph Zechner for stimulating discussions. GT and SACH were funded in part by Austrian Science Fund grant FWF P28844.

\clearpage 
\pagebreak

\section*{Supplementary Figures}

\renewcommand{\thefigure}{S\arabic{figure}}

\setcounter{figure}{0}

\begin{figure}[H] 
	\centering
	\includegraphics[width=.7\textwidth]{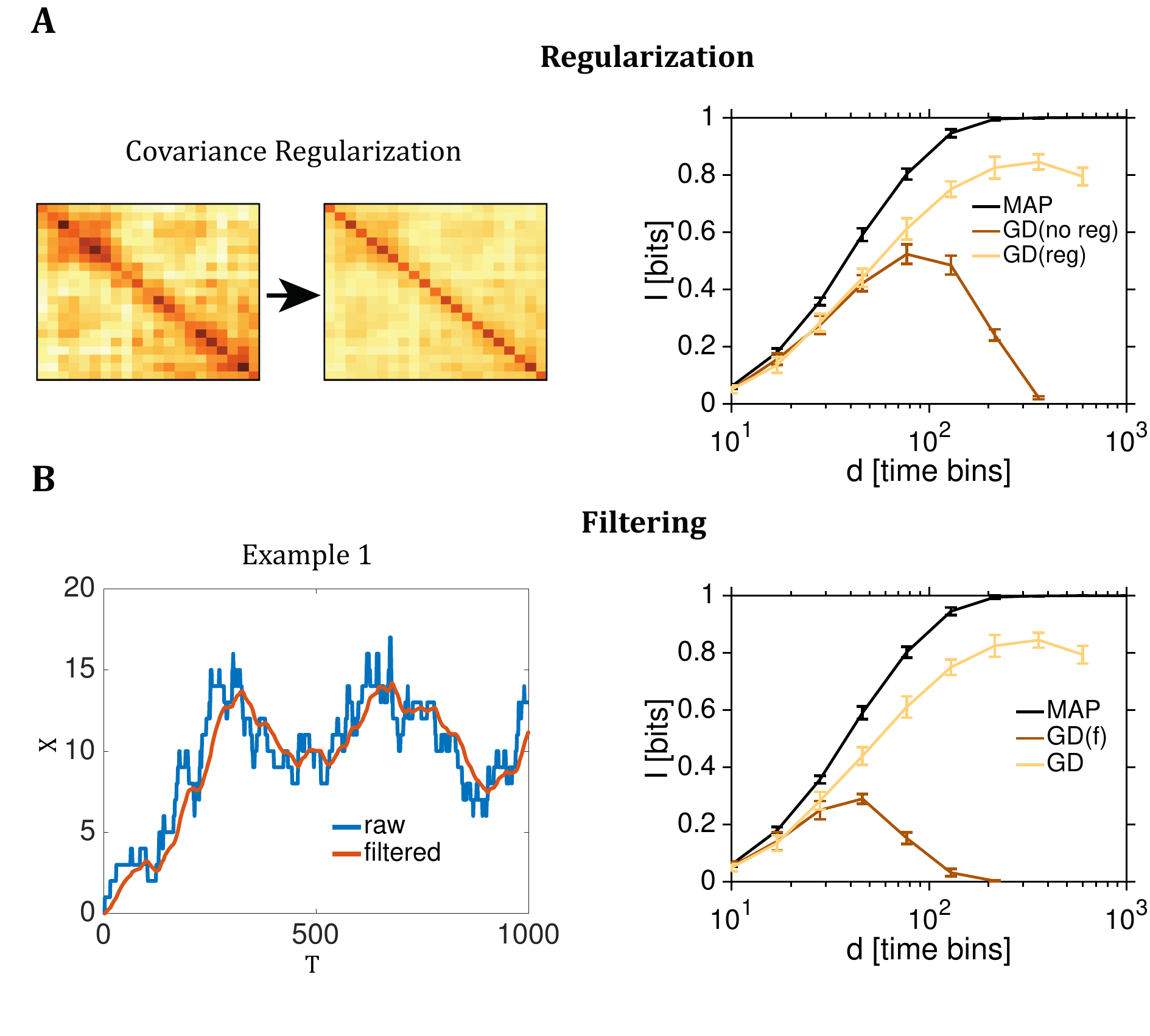}	
	\caption{{\bf Effects of covariance matrix regularization and signal smoothing on Gaussian-decoder-based estimation. } {\bf (A)} At left. Diagonal covariance regularization following  Ref \cite{Yatsenko2015}. Briefly, $\lambda$ times the identity matrix is added to the empirical covariance matrix with the hyperparameter $\lambda$ set so that the likelihood on test data is maximized. Shown is the empirical (left) and regularized (right) covariance matrix for Example 3, using $d=20$ and $N=30$ sample trajectories. 
	 At right. Information estimates for Example 3: $I_{\rm MAP}$ decoding bound (black), Gaussian decoder estimate, $I_{\rm GD(reg)}$, with optimal diagonal regularization for each $d$ (yellow, as in Fig~\ref{fig:CasesInfoDecodeNtb}C), Gaussian decoder estimate, $I_{\rm GD(no reg)}$ (brown). Without regularization, the estimate suffers an abrupt drop as $d$ increases and the empirically estimated covariance matrix becomes close to singular. $N$ and plotting conventions are as in Fig~\ref{fig:CasesInfoDecodeNtb}. 
	{\bf (B)} The effects of trajectory filtering on information estimates. At left. A raw integer-valued stochastic trajectory for $\tilde{X}$ (blue) can be filtered by a low-pass exponential decay filter with adjustable timescale, $\tau = 1-10^3$, here $\tau=50$ (red) to yield real-valued trajectory. At right. Regularized Gaussian-decoder information estimates with (brown) and without (yellow) filtering. Filtering does not improve but can decrease the estimation performance, even when the filtering timescale is adjusted.}	\label{fig:CS3GDReg}	
\end{figure}

\begin{figure}[H] 
	\centering
	\includegraphics[width=.35\textwidth]{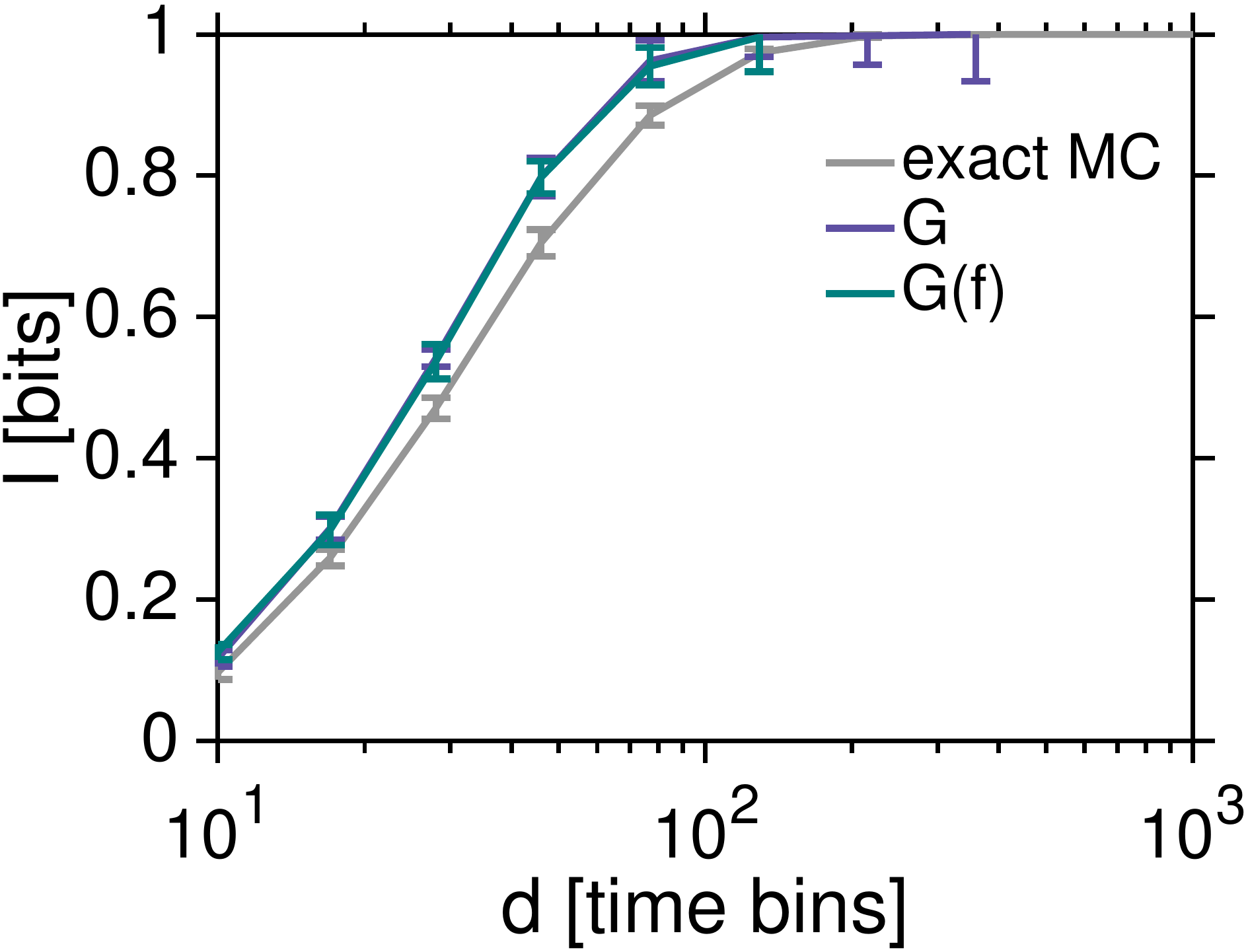}
	\caption{{\bf Gaussian approximation to the information can lead to an uncontrolled overestimation of the true information.} Gaussian approximation is evaluated for Example 3 in Fig~\ref{fig:CasesInfoDecodeNtb}C, using  $N=1000$ per condition. Exact Monte Carlo approximation  of the information, $I_{\rm exact}(\mathbf{X};U)$, is shown in dark gray. Information estimates following Methods Section~\ref{sec:mfe} are shown in violet (Gaussian approximation for raw, integer-valued response trajectories) or in cyan (Gaussian approximation for filtered trajectories, as in Fig~\ref{fig:CS3GDReg}). In both cases the Gaussian approximation overshoots the true information value. Further numerical analyses (not shown) indicate that the difference is hard to predict and that it persists even when the reaction rates are chosen such that the mean expression level is ten-fold higher (and the intrinsic stochasticity correspondingly lower). This makes direct Gaussian approximation risky to use, in contrast to the Gaussian-decoder based estimate, which is guaranteed to stay below $I_{\rm exact}$. }
	\label{fig:CS3GaussFiltering}	
\end{figure}

\begin{figure}[H] %
	\centering
	\includegraphics[width=.45\textwidth]{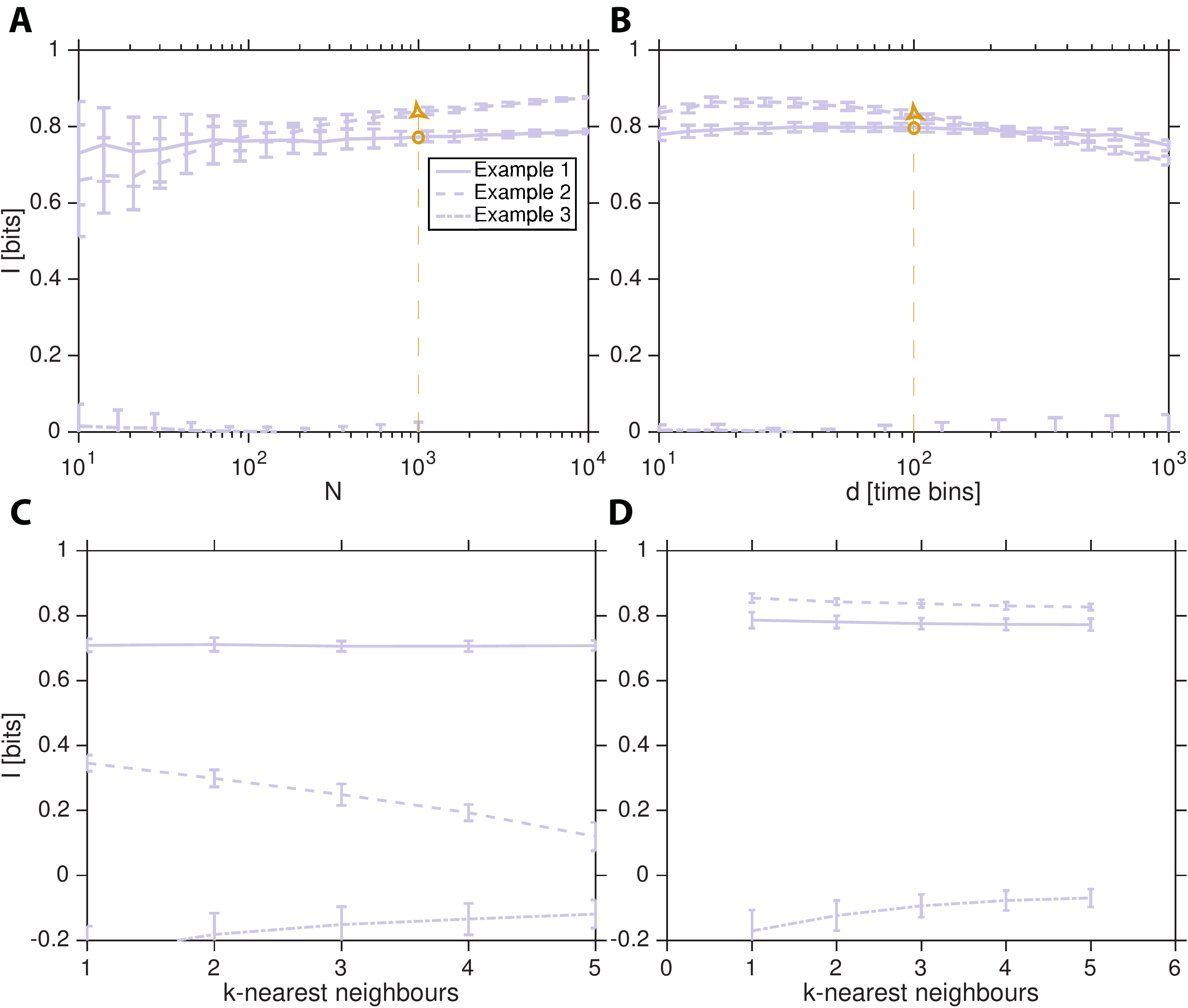}
	\caption{{\bf Behavior of the knn information estimator.} Compared to knn results in Fig~\ref{fig:BarplotCases}, the results in A, B and D are estimated following the same procedure, while adding a small amount of IID zero-mean Gaussian noise to each response trajectory at every time bin; the noise variance must be $\ll 1$ but otherwise does not affect the results much. This results in good estimates even at low sample number, $N$, and provides nearly stable estimation as a function of the trajectory dimension, $d$, for Example 1 and Example 2. It, however, does not resolve the estimator failure for Example 3. {\bf (A)} Dependence of the knn estimator performance on the number of samples. Yellow plot symbols indicate the number of samples per condition, $N=10^3$, used in Fig~\ref{fig:BarplotCases}. {\bf (B)} Dependence of the knn estimator performance on the trajectory dimension. Yellow plot symbols indicate the dimension, $d=10^2$, used in Fig~\ref{fig:BarplotCases}. {\bf (C, D)} Dependence of the knn estimates on the number of nearest neighbors, $k$, at $N=10^3$ and $d=10^2$, without the addition of noise (C) or with the addition of noise (D).}
	\label{fig:knn_cs123}	
\end{figure}

\begin{figure}[H] 
	\centering
	\includegraphics[width=.35\textwidth]{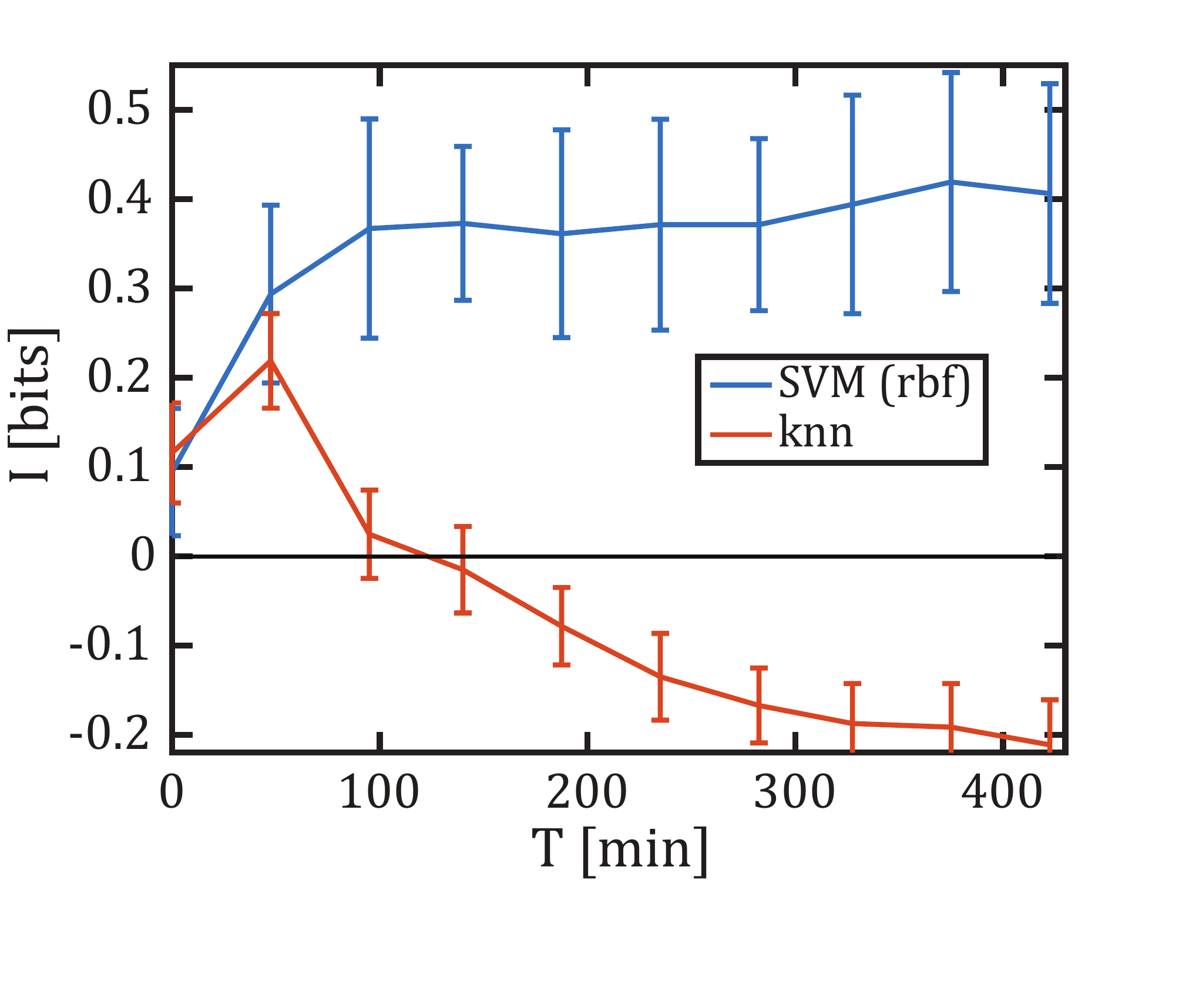}	
	\caption{{\bf Estimator behavior for longer trajectory data for Dot6.} When the samples are limited, here to $N=100$ samples per input glucose level condition as in Fig~\ref{fig:InfoEstData}A (middle), radial-basis-function SVM estimate (blue) is  well-behaved with no observable overfitting and consequent drop in information estimate as the trajectory duration, $T$, is increased (maximal $T$ corresponds to $d=170$ dimensional trajectory vectors). In contrast, knn estimate (brown) shows a collapse in the estimation performance, even yielding strongly negative numbers, as the dimensionality of input vectors is increased at fixed number of trajectory samples. } 
	\label{fig:knn2}	
\end{figure}

\begin{figure}[H] %
	\centering
	\includegraphics[width=.4\textwidth]{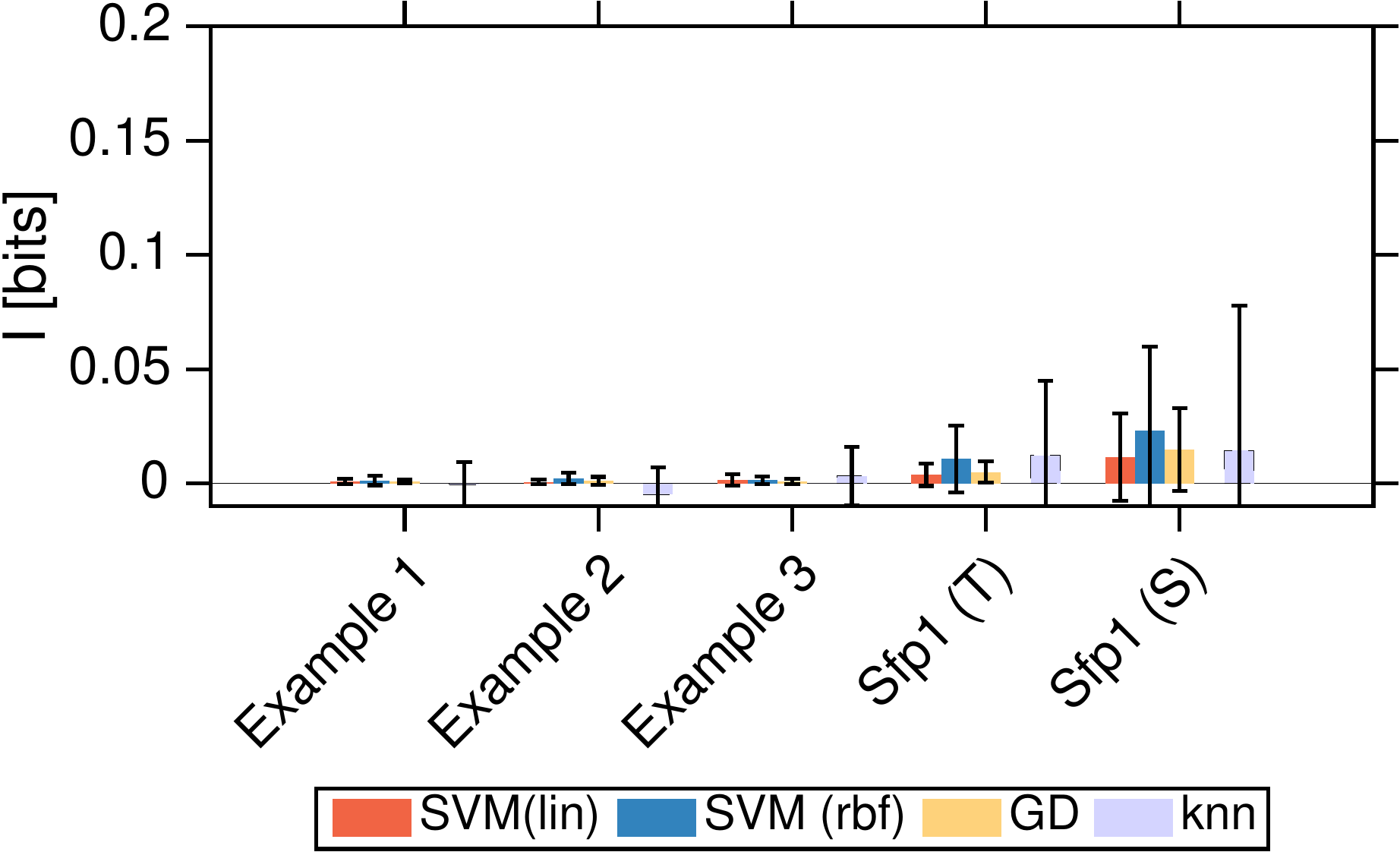}
	\caption{{\bf Assessing information estimation bias due to small sample size.} By randomly shuffling the binary labels assigned to different response trajectories, we break all response-input correlations leading to zero information. Here we test whether our estimators correctly report zero information within error bars given a finite number of samples, or are subject to positive information estimation bias. Decoding-based estimates (linear SVM, red; kernelized SVM, blue; Gaussian decoder, yellow) and knn (gray). First three sets of bars correspond to synthetic examples of Fig~\ref{fig:InfoCont}; estimations are done with $d=100$ and $N=1000$ per input condition as in Figs~\ref{fig:CasesInfoDecodeNtb} and \ref{fig:CasesInfoDecodeN}, following the same plotting conventions. Last two sets of bars are estimated with $N=100$ per input condition using real data for Sfp1 yeast TF from Fig~\ref{fig:InfoEstData}A. In all cases, even without explicit small-sample debiasing for Eq~(\ref{eq:errorinfo}) (which may be required for multilevel estimation), the estimates are consistent with zero. }
	\label{fig:RandLabeling}	
\end{figure}

\begin{figure}[H] %
	\centering
	\includegraphics[width=.7\textwidth]{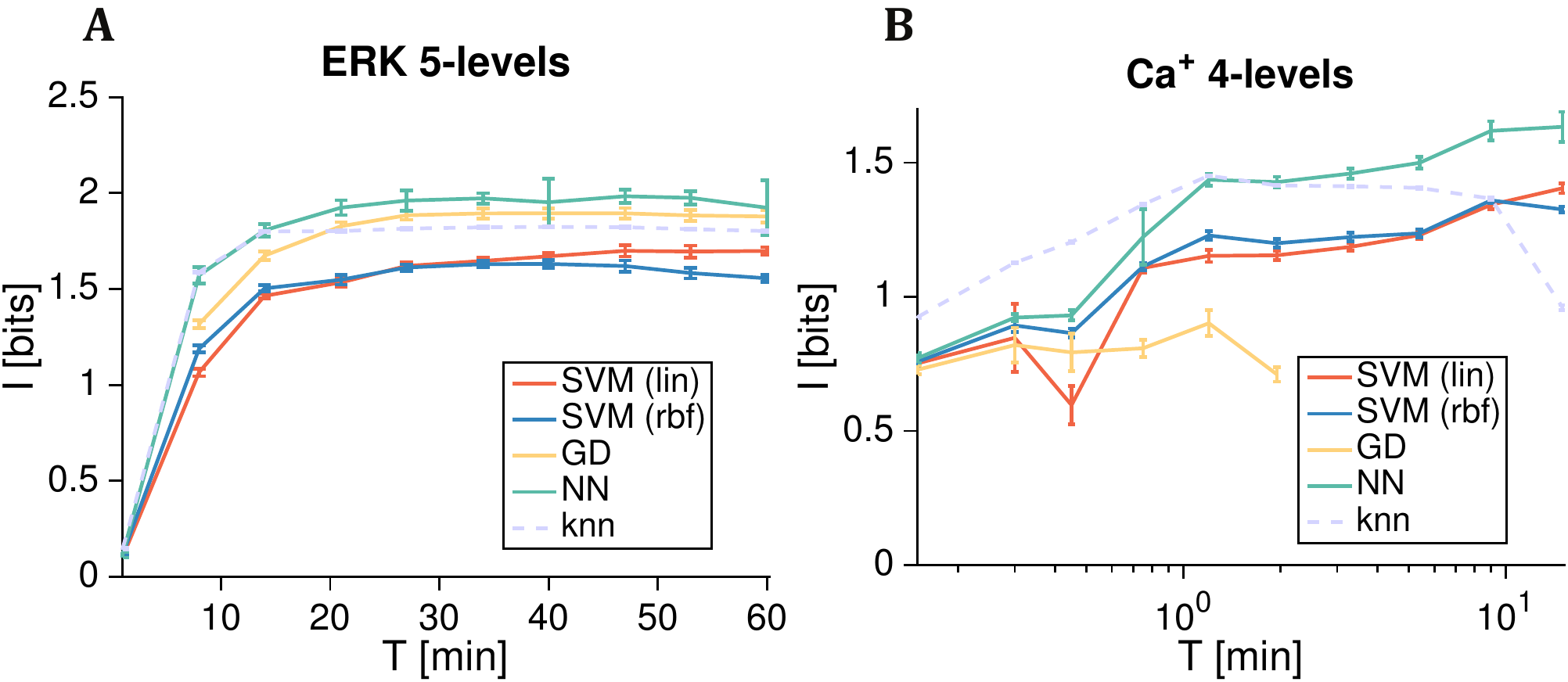}
	\caption{{\bf Information estimates for mammalian signaling networks as a function of the trajectory duration.}  Shown are information estimates as a function of the total trajectory duration, $T$, for the early response period for ERK (left) and Ca$^+$ (right). Plotting conventions, procedures, and data set sizes same as in Fig~\ref{fig:InfoEstDataERK_CA}. } 
	\label{fig:ERK_CA}	
\end{figure}

\clearpage
\pagebreak

\bibliographystyle{unsrt}
\bibliography{Project2Theory.bib,Proy2_a.bib}

\end{document}